\UseRawInputEncoding
\documentclass[showpacs,aps,graphicx,twocolumn]{revtex4}%and
\usepackage{graphicx}
\pdfoutput=1

\begin{document}
\title{Advances in quantum entanglement purification}

\author{Pei-Shun Yan$^{1,2,4}$, Lan Zhou$^{3}$, Wei Zhong$^{4}$, and Yu-Bo Sheng$^{1,4}$\footnote{Email address:
shengyb@njupt.edu.cn}  }
\address{
$^1$College of Electronic and Optical Engineering \& College of Flexible Electronics (Future Technology), Nanjing University of Posts and Telecommunications, 210023, Nanjing, China\\
$^2$School of Information Science and Technology, Nantong University, 226019, Nantong, China\\
$^3$School of Science, Nanjing University of Posts and Telecommunications, 210023, Nanjing, China\\
$^4$Institute of Quantum Information and Technology, Nanjing University of Posts and Telecommunications, 210003, Nanjing, China\\
}

\begin{abstract}
Since its discovery, the quantum entanglement becomes a promising resource in quantum communication and computation. However, the entanglement is fragile due to the presence of noise in quantum channels. Entanglement purification is a powerful tool to distill high quality entangled states from the low quality entangled states. In this review, we present an overview of entanglement purification, including the basic entanglement purification theory, the entanglement purification protocols (EPPs) with linear optics, EPPs with cross-Kerr nonlinearities, hyperentanglement EPPs, deterministic EPPs, and measurement-based EPPs. We also review experimental progresses of EPPs in linear optics. Finally, we give the discussion on potential outlook for the future development of EPPs. This review may pave the way for practical implementations in future long-distance quantum communication and quantum network.


\end{abstract}

\maketitle

\section{Introduction}
Quantum entanglement is a counterintuitive phenomenon which leads quantum mechanics to be different from classical one and it is a central resource in quantum information processing such as quantum key distribution (QKD) \cite{qkd1,qkd2,qkd3,qkd4,qkd5}, quantum secret sharing (QSS) \cite{qss1,qss2}, quantum secure direct communication (QSDC) \cite{qsdc1,qsdc2,qsdc4,qsdc5,qsdc51,qsdc6,qsdc61,qsdc7,qsdc8,qsdc81,qsdc10,qsdc11,qsdc12,qsdc13,qsdc14,qsdc15}, quantum teleportation \cite{QT1,QT2,QT3,QT4}, quantum computation \cite{QuantumComputation1,QuantumComputation2,QuantumComputation3,QuantumComputation4}, and quantum network \cite{quantumnetwork1,quantumnetwork2,quantumnetwork3,quantumnetwork4,quantumnetwork5}.  However, the inherent noise of quantum channels and the imperfect operations will lead to the maximal entangled states degrade into the low quality mixed states \cite{repeater1}. All of these defects limit the applications of quantum entanglement and  degrade the performance of quantum information processing.

To obtain the high-quality entanglement, one can use some methods such as the high-fidelity entanglement generation \cite{EG1,EG3,EG4,EG6,EG7}, quantum error correction codes (QECCs) \cite{errorcorrection1,errorcorrection2,errorcorrection3,errorcorrection4,errorcorrection5,errorcorrection6,errorcorrection7,errorcorrection8}, and entanglement purification protocols (EPPs) \cite{EPP1,EPP2,EPP3,EPP4,EPP5,EPP51,EPP27,EPP6,EPP7,EPP8,EPP9,EPP28,EPP10,WangOE1,EPP12,EPP13,
EPP14,EPP15,EPP18,EPP16,EPP17,EPP18Science,EPP19,EPP20,EPP21,EPP22,EPP23,EPP24,EPP25,EPPaddyan,EPPPRAyan,EAEPP1,EAEPP2,EPPSB,EPPadd2,EPPPRL1,
HighErrorTolerance1,HighErrorTolerance2,EPPLuo}. However, the imperfections in physical devices will reduce the quality of entanglement. It is still fragile during entanglement distribution over noisy channels. Additionally, the QECC is generally adopted to prevent unknown states from being damaged \cite{EAEPP1,EAEPP2}. Moreover, the requirements of the QECC are strict and the error tolerance threshold of this method is rather low \cite{EAEPP1,EAEPP2}. By contrast, the target entangled state of EPPs is known \cite{EAEPP1,EAEPP2} and the error tolerance threshold is higher than that of QECC \cite{HighErrorTolerance1,HighErrorTolerance2}.
Furthermore, the EPP is a powerful tool to distill fewer high-quality entangled states from a large number of less-quality copies with local operations and classical communication \cite{EPP1}, which is a key role in quantum repeaters because it determines the communication efficiency in long-distance quantum communication \cite{repeater1,EPPinQR1,EPPinQR2,EPPinQR3,EPPinQR4,EPPinQKD1} and quantum network \cite{EPPinQN1,EPPinQN2,EPPinQN3}.

In this review paper, we focus on the  development of EPPs. We review the conventional EPPs based on the controlled-not (CNOT) gates or similar logical operations for bipartite systems in Sec. II. Then, we introduce some EPPs for multipartite systems in Sec. III. In \textbf{Sec. IV}, the hyperentanglement EPPs including purification for the polarization degree of freedom (DOF) using other DOFs and the EPP for nonlocal hyperentangled systems will be introduced. In Sec. V, we introduce a novel EPP named measurement-based EPP (MBEPP). In Sec. VI, we discuss some possible future development of EPPs.

%Additionally, the feasibility of this MBEPP in linear optics is highlighted using spontaneous parametric down conversion (SPDC) sources.

\section{EPPs based on the CNOT gates or similar logical operations for bipartite systems}
In this section, we mainly focus on the EPPs based on the CNOT gates or similar logical operations for bipartite systems. The first EPP was proposed by Bennett \emph{et al.} \cite{EPP1}, in which the CNOT gates are acted on two identical Werner states. After performing the CNOT operations, the source pair is retained if the measurement outcomes of the target pair are the same. Otherwise, one can discard the source pair. In this way, the high-quality entangled state is obtained provided that the initial fidelity is larger than 0.5. Sequently, Deutsch \emph{et al.}
extended the EPP to the case of arbitrary mixed states thereby improving efficiency of the EPP \cite{EPP2}. In 2001, Pan \emph{et al.} proposed an efficient EPP to purify mixed states with ideal sources in linear optics \cite{EPP3}. In 2003, Pan \emph{et al.} performed the first entanglement purification experiment using spontaneous parametric down conversion (SPDC) sources \cite{EPP5}. In 2008, Maruyama \emph{et al.} used two-spin operations for an isotropic Heisenberg interaction to realize CNOT gate to purify the polluted entanglement \cite{EPP27}. In the same year, Sheng \emph{et al.} proposed an efficient polarization-entanglement purification protocol based on the SPDC sources with the cross-kerr nonlinearity \cite{EPP6}. In 2011, Wang \emph{et al.} proposed an EPP to purify the electron-spin entanglement \cite{EPP9}. Subsequently, they investigated the hybrid-EPP in the coupling systems \cite{EPP28}. In 2013, Sheng \emph{et al.} proposed the hybrid-EPP for the hybrid entanglement in linear optics \cite{EPP10}. In 2015, Wang \emph{et al.} resorted a parity-checking and qubit amplifying (PCQA) gate in linear optics to against the transmission loss and the decoherence in a high-efficient method \cite{WangOE1}. In 2017, Zhang \emph{et al.} investigated an EPP for nonlocal microwave photons by employing the cross-Kerr effect in circuit quantum electrodynamics \cite{EPP18}. In the same year, a nested EPP for quantum repeater was experimentally reported, which not only enlarges the communication distance but also eliminates the double-pair emissions from the SPDC sources \cite{EPP16}. In Ref. \cite{EPP19}, an efficient EPP for $d$-level systems was proposed and the robustness and efficiency of the EPP will be improved with an increased dimension. In 2020, Zhou \emph{et al.} discussed the EPP with non-identical mixed states \cite{EPP21}. They found that the discarding components in the conventional EPPs \cite{EPP1,EPP2} may have residual entanglement and can be reused in a next round, which increases the efficiency of entanglement purifications.

In  this section, we mainly introduce the original EPP based on the CNOT gates \cite{EPP1} and the EPP based on the polarization beam splitters (PBSs) \cite{EPP3}. Additionally, we will also review experimental EPPs in linear optics \cite{EPP5,EPP16}.

\subsection{The EPP based on the CNOT gates}
We describe the first EPP based on the CNOT gates \cite{EPP1}. Suppose that Alice and Bob intend to share the state $|\phi_n^+\rangle_{ab}$, which is one of the four Bell states
\begin{eqnarray}\label{Bellstates}
{|{{\phi_n^\pm }}\rangle _{ab}} = \frac{1}{{\sqrt 2 }}({| 0\rangle _a}{| 0\rangle _b} \pm {| 1\rangle _a}{| 1\rangle _b}),\nonumber\\
{| {{\psi_n^ \pm }}\rangle _{ab}} = \frac{1}{{\sqrt 2 }}({| 0\rangle _a}{| 1\rangle _b} \pm {| 1\rangle _a}{| 0 \rangle _b}).
\end{eqnarray}
However, the maximally entangled state will collapse to the mixed state due to the inherent noise in quantum channels, yielding
\begin{eqnarray}\label{CNOTmixed}
\rho_{ab}  &=& F{|{{\phi_n^+}}\rangle_{ab}}\langle {{\phi_n^+}}| + A{| {{\phi_n^-}}\rangle_{ab}}\langle{{\phi_n^-}}|\nonumber\\
&+&B{| {{\psi_n^+}}\rangle_{ab}}\langle{{\psi_n^+}}|+C{| {{\psi_n^-}}\rangle_{ab}}\langle{{\psi_n^-}}|,
\end{eqnarray}
where $F+A+B+C=1$. Any mixed state likes Eq. (\ref{CNOTmixed}) can be transformed to the Werner state by the combinations of bilateral rotations and unilateral rotations. Thus, we can obtain
\begin{eqnarray}\label{Wernerstate}
{\rho _{ab}^{\prime}} &=& F|{\phi_n^ + }{\rangle _{ab}}\langle {\phi_n^ + }| + \frac{{1 - F}}{3}{(|{\phi_n^ - }\rangle _{ab}}\langle {\phi_n^ - }|\nonumber\\
 &+& |{\psi_n^ + }{\rangle _{ab}}{\langle {\psi_n^ + }| + |{\psi_n^ - }\rangle _{ab}}\langle {\psi_n^ - }|).
\end{eqnarray}
As shown in Fig. 1, two noisy copies $\rho_{{a_1}{b_1}}$ and $\rho_{{a_2}{b_2}}$ with the same form as Eq. (\ref{Wernerstate}) are required to perform the purification. The whole system $\rho_{{a_1}{b_1}}\otimes\rho_{{a_2}{b_2}}$ can be described as follows. It is in the state $|\phi_n^+\rangle_{{a_1}{b_1}}\otimes|\phi_n^+\rangle_{{a_2}{b_2}}$ with the probability of $F^2$. With an equal probability of $\frac{{{{(1 - F)}^2}}}{9}$, the system is in $|\phi_n^-\rangle_{{a_1}{b_1}}\otimes|\phi_n^-\rangle_{{a_2}{b_2}}$, $|\psi_n^+\rangle_{{a_1}{b_1}}\otimes|\psi_n^+\rangle_{{a_2}{b_2}}$, and $|\psi_n^-\rangle_{{a_1}{b_1}}\otimes|\psi_n^-\rangle_{{a_2}{b_2}}$. For the other cases, the whole system is in the cross combinations with a specific probability.

\begin{figure}[!h]
\begin{center}
\includegraphics[scale=2.1]{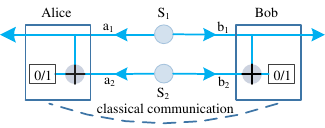}
\caption{The schematic diagram of the EPP based on the CNOT gates \cite{EPP1}. Two sources $S_1$ and $S_2$ generate two pairs of entanglement. We assume that $a_1b_1$ ($a_2b_2$) is control pair (target pair). After the CNOT gates, if the control qubit is $|0\rangle$, the target qubit remains unchanged. If the control qubit is $|1\rangle$, it is essential to perform bit-flip operation on the target qubit, i.e., $|0\rangle\leftrightarrow|1\rangle$. Subsequently, the target pair is measured with the Z-basis and the source pair will be retained as long as the same measurement outcome is obtained. Otherwise, we discard the source pair.}
\end{center}
\end{figure}
As depicted in Fig. 1, the photon pair ${a_1b_1}$ is the source pair and ${a_2b_2}$ is the target pair. After passing through the CNOT gates, the target qubit remains unchanged if the control qubit is $|0\rangle$. If the control qubit is $|1\rangle$, the bit-flip operation should be taken place on the target qubit, i.e., $|0\rangle\leftrightarrow|1\rangle$. Thus, the initial photon pairs $a_1b_1$ and $a_2b_2$ will become
\begin{eqnarray}\label{phiphi1}
&&|\phi _n^ \pm {\rangle _{{a_1}{b_1}}} \otimes |\phi _n^ + {\rangle _{{a_2}{b_2}}} \to |\phi _n^ \pm {\rangle _{{a_1}{b_1}}} \otimes |\phi _n^ + {\rangle _{{a_2}{b_2}}},\nonumber\\
&&|\phi _n^ \pm {\rangle _{{a_1}{b_1}}} \otimes |\phi _n^ - {\rangle _{{a_2}{b_2}}} \to |\phi _n^ \mp {\rangle _{{a_1}{b_1}}} \otimes |\phi _n^ - {\rangle _{{a_2}{b_2}}},\nonumber\\
&&|\phi _n^ \pm {\rangle _{{a_1}{b_1}}} \otimes |\psi _n^ + {\rangle _{{a_2}{b_2}}} \to |\phi _n^ \pm {\rangle _{{a_1}{b_1}}} \otimes |\psi _n^ + {\rangle _{{a_2}{b_2}}},\nonumber\\
&&|\phi _n^ \pm {\rangle _{{a_1}{b_1}}} \otimes |\psi _n^ - {\rangle _{{a_2}{b_2}}} \to |\phi _n^ \mp {\rangle _{{a_1}{b_1}}} \otimes |\psi _n^ - {\rangle _{{a_2}{b_2}}},\nonumber\\
&&|\psi _n^ \pm {\rangle _{{a_1}{b_1}}} \otimes |\phi _n^ + {\rangle _{{a_2}{b_2}}} \to |\psi _n^ \pm {\rangle _{{a_1}{b_1}}} \otimes |\psi _n^ + {\rangle _{{a_2}{b_2}}},\nonumber\\
&&|\psi _n^ \pm {\rangle _{{a_1}{b_1}}} \otimes |\phi _n^ - {\rangle _{{a_2}{b_2}}} \to |\psi _n^ \mp {\rangle _{{a_1}{b_1}}} \otimes |\psi _n^ - {\rangle _{{a_2}{b_2}}},\nonumber\\
&&|\psi _n^ \pm {\rangle _{{a_1}{b_1}}} \otimes |\psi _n^ + {\rangle _{{a_2}{b_2}}} \to |\psi _n^ \pm {\rangle _{{a_1}{b_1}}} \otimes |\phi _n^ + {\rangle _{{a_2}{b_2}}},\nonumber\\
&&|\psi _n^ \pm {\rangle _{{a_1}{b_1}}} \otimes |\psi _n^ - {\rangle _{{a_2}{b_2}}} \to |\psi _n^ \mp {\rangle _{{a_1}{b_1}}} \otimes |\phi _n^ - {\rangle _{{a_2}{b_2}}}.
\end{eqnarray}

Then, one can measure the target pair with the Z-basis. If the measurement outcomes of Alice and Bob are the same, it indicates a successful purification. Therefore, we retain the source pair. Otherwise, the source pair will be discarded. Hence, the new mixed state can be written as
\begin{eqnarray}\label{CNOTnewmixed}
{\rho _{ab}^{\prime\prime}} &=& {A_1}|{\phi_n ^ + }{\rangle _{ab}}{\langle {\phi_n ^ + }| + {B_1}|{\phi _n^ - }\rangle _{ab}}\langle {\phi_n ^ - }|\nonumber\\
 &+& {C_1}(|{\psi_n ^ + }{\rangle _{ab}}{\langle {\psi _n^ + }| +|{\psi_n ^ - }\rangle _{ab}}\langle {\psi_n ^ - }|),
\end{eqnarray}
where ${A_1} = \frac{1}{N}[{F^2} + \frac{1}{9}{(1 - F)^2}]$, ${B_1} = \frac{2}{{3N}}F(1 - F)$,${C_1} = \frac{2}{{9N}}{(1 - F)^2}$ and $N = {F^2} + \frac{5}{9}{(1 - F)^2} + \frac{2}{{3N}}F(1 - F)$. If $F>0.5$, the fidelity of the resultant state is higher than that of initial one. After one round of this EPP, the rate of bit-flip errors is reduced simultaneously improving the rate of phase-flip errors. To address this issue, bilateral and unilateral rotations should be acted on qubits to transform the state as Eq. (\ref{CNOTnewmixed}) to the Werner state before each round of purification. As a result, the efficiency of the EPP based on the CNOT gates is rather low \cite{EPP1}. To this end, Deutch \emph{et al.} added unitary operations  \cite{EPP2} to the input state such as
\begin{eqnarray}\label{Deutch}
&&|0{\rangle _a} \to \frac{1}{{\sqrt 2 }}{(|0\rangle _a} - i|1{\rangle _a}),\nonumber\\
&&|1{\rangle _a} \to \frac{1}{{\sqrt 2 }}{(|1\rangle _a} - i|0{\rangle _a}),\nonumber\\
&&|0{\rangle _b} \to \frac{1}{{\sqrt 2 }}{(|0\rangle _b} + i|1{\rangle _b}),\nonumber\\
&&|1{\rangle _b} \to \frac{1}{{\sqrt 2 }}{(|1\rangle _b} + i|0{\rangle _b}).
\end{eqnarray}
Subsequently, the next step as Ref. \cite{EPP1} can be adopted. In this way, this EPP enables to purify the arbitrary mixed states and the efficiency of the EPP can be significantly enhanced.

\begin{figure}[!h]
\begin{center}
\includegraphics[scale=2.1]{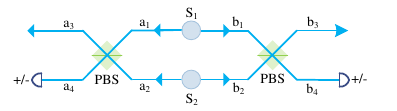}
\caption{The schematic diagram of the EPP based on the PBSs \cite{EPP3}. The PBS totally transmits the photon in polarization $|H\rangle$ and reflects the photon in polarization $|V\rangle$. The success purification requires each output modes $a_3$, $b_3$, $a_4$, and $b_4$ precisely contains one photon, which is named ``four mode'' case. Then, one needs to measure the photons in modes $a_4$ and $b_4$ with the basis $\{ | + \rangle ,| - \rangle \}$, where $| \pm \rangle  = \frac{1}{{\sqrt 2 }}(|H\rangle  \pm |V\rangle )$. Finally, the fidelity of mixed state can be increased.}
\end{center}
\end{figure}

\subsection{The EPP based on the PBSs}
It is known that a deterministic CNOT gate is hard to be realized in linear optics, which seems to an obstacle for entanglement purification. To solve this problem, the first EPP in linear optics was proposed in 2001 \cite{EPP3}, which used the PBS to play a role of the CNOT gate between the polarization and the spatial mode. Consequently, one can merely employ the PBSs to carry out the purification instead of the CNOT gates. This EPP requires each of output modes precisely contains one photon, named ``four mode'' case, to herald a successful purification. Let's suppose that the initial mixed state is
\begin{eqnarray}\label{PBSmixed}
\rho_{ab}^{{\prime}{\prime}{\prime}} = F{| {{\phi ^ + }} \rangle _{ab}}\langle {{\phi ^ + }} | + (1 - F){| {{\psi ^ + }}\rangle _{ab}}\langle {{\psi ^ + }}|.
\end{eqnarray}
where
\begin{eqnarray}\label{PBS}
|{\phi ^ \pm }{\rangle _{ab}} = \frac{1}{{\sqrt 2 }}{(|H\rangle _a}|H{\rangle _b} \pm |V{\rangle _a}|V{\rangle _b}),\nonumber\\
|{\psi ^ \pm }{\rangle _{ab}} = \frac{1}{{\sqrt 2 }}{(|H\rangle _a}|V{\rangle _b} \pm |V{\rangle _a}|H{\rangle _b}).
\end{eqnarray}
Here $|H\rangle$ and $|V\rangle$ respectively denote the horizontal polarization and the vertical polarization of a photon.
 As shown in Fig. 2, $\rho_{a_1b_1}^{{\prime}{\prime}{\prime}} \otimes\rho_{a_2b_2}^{{\prime}{\prime}{\prime}} $ is made up of four pure states such as $|\phi^+\rangle_{a_1b_1}\otimes|\phi^+\rangle_{a_2b_2}$, $|\phi^+\rangle_{a_1b_1}\otimes|\psi^+\rangle_{a_2b_2}$, $|\psi^+\rangle_{a_1b_1}\otimes|\phi^+\rangle_{a_2b_2}$, and $|\psi^+\rangle_{a_1b_1}\otimes|\psi^+\rangle_{a_2b_2}$ with the probability of $F^2$, $F(1-F)$, $(1-F)F$, and $(1-F)^2$, separately. After the PBSs, the state $|\phi^+\rangle_{a_1b_1}\otimes|\phi^+\rangle_{a_2b_2}$ evolves to
\begin{eqnarray}\label{PBSevolved}
&&|{\phi ^ + }{\rangle _{{a_1}{b_1}}} \otimes |{\phi ^ + }{\rangle _{{a_2}{b_2}}}\nonumber\\
&\to&\frac{1}{2}{(|H\rangle _{{a_3}}}|H{\rangle _{{b_3}}}|H{\rangle _{{a_4}}}|H{\rangle _{{b_4}}} \!+\! |H{\rangle _{{a_4}}}|H{\rangle _{{b_4}}}|V{\rangle _{{a_4}}}|V{\rangle _{{b_4}}}\nonumber\\
&+&|V{\rangle _{{a_3}}}|V{\rangle _{{b_3}}}|H{\rangle _{{a_3}}}|H{\rangle _{{b_3}}} \!+\! |V{\rangle _{{a_3}}}|V{\rangle _{{b_3}}}|V{\rangle _{{a_4}}}|V{\rangle _{{b_4}}}).
\end{eqnarray}
Obviously, the two items $|H{\rangle _{{a_3}}}|H{\rangle _{{b_3}}}|H{\rangle _{{a_4}}}|H{\rangle _{{b_4}}}$ and $|V{\rangle_{{a_3}}}|V{\rangle _{{b_3}}}|V{\rangle _{{a_4}}}|V{\rangle _{{b_4}}}$ make each of output modes contain one photon. While the other components $|H{\rangle _{{a_3}}}|H{\rangle _{{b_3}}}|V{\rangle _{{a_4}}}|V{\rangle _{{b_4}}}$ and  $|V{\rangle _{{a_3}}}|V{\rangle _{{b_3}}}|H{\rangle _{{a_4}}}|H{\rangle _{{b_4}}}$ fail to satisfy the ``four mode'' case. Thus, they can be washed out automatically. Then, we utilize the basis $\{ | + \rangle ,| - \rangle \}$ to measure the photons in modes $a_4$ and $b_4$. If the measurement results are $|+\rangle|+\rangle$ or $|-\rangle|-\rangle$, the resultant state is $|\phi^+\rangle_{a_3b_3}$. If the measurement results are $|+\rangle|-\rangle$ or $|-\rangle|+\rangle$, the resultant state is $|\phi^-\rangle_{a_3b_3}$. In this case, an additional phase-flip operation ${\sigma _z} = |H\rangle\langle H| - |V\rangle\langle V|$ should be performed on one of two photons. With the same principle, the state $|\psi^+\rangle_{a_1b_1}\otimes|\psi^+\rangle_{a_2b_2}$ collapses to $|\psi^+\rangle_{a_3b_3}$ after measuring the photons in modes $a_4$ and $b_4$. However, the cross combinations $|\phi^+\rangle_{a_1b_1}\otimes|\psi^+\rangle_{a_2b_2}$ and $|\psi^+\rangle_{a_1b_1}\otimes|\phi^+\rangle_{a_2b_2}$ can be removed automatically because they only satisfy the ``three-mode'' case. For instance,
\begin{eqnarray}\label{crosscombination}
&&|{\phi ^ + }{\rangle _{{a_1}{b_1}}} \otimes |{\psi ^ + }{\rangle _{{a_2}{b_2}}}\nonumber\\
 \!&\to&\! \frac{1}{2}{(|H\rangle _{{a_3}}}|H{\rangle _{{a_4}}}|H{\rangle _{{b_4}}}|V{\rangle _{{b_4}}} \!+\! |H{\rangle _{{a_4}}}|V{\rangle _{{a_4}}}|H{\rangle _{{b_4}}}|H{\rangle _{{b_3}}}\nonumber\\
\!&+&\!|H{\rangle _{{a_3}}}|V{\rangle _{{a_3}}}|V{\rangle _{{b_3}}}|V{\rangle _{{b_4}}} \!+\! |V{\rangle _{{a_3}}}|V{\rangle _{{a_4}}}|V{\rangle _{{b_3}}}|H{\rangle _{{b_3}}}).\nonumber\\
\end{eqnarray}
which shows that the items of Eq. (\ref{crosscombination}) lead to the coincidence detections on $a_3a_4b_4$ or $b_3a_4b_4$ or $a_3b_3b_4$ or $a_3b_3a_4$, indicating the ``three-mode'' case. Hence, it can be automatically eliminated. Consequently, we can obtain a new mixed state written as
\begin{eqnarray}\label{PBSnewmixed}
\rho_{a_3b_3} = F_1{| {{\phi ^ + }} \rangle _{a_3b_3}}\langle {{\phi ^ + }} | + (1-F_1){| {{\psi ^ + }}\rangle _{a_3b_3}}\langle {{\psi ^ + }}|,
\end{eqnarray}
with the fidelity $F_1=\frac{{{F^2}}}{{{F^2} + {{(1 - F)}^2}}}$. The success probability of this EPP is $\frac{{{F^2} + {{(1 - F)}^2}}}{2}$.

\begin{figure}
\begin{center}
\includegraphics[scale=2.1]{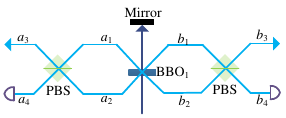}
\caption{The schematic diagram of the EPP based on the SPDC source in linear optics \cite{EPP5}. The BBO means the beta barium borate.}
\end{center}
\end{figure}

The EPP in Ref. \cite{EPP3} is designed for the ideal entanglement sources. However, the current available entanglement source such as the SPDC source works in a probability way and it seems that the SPDC source is unsuitable for entanglement purification. Interestingly, the SPDC source is not an obstacle to be implemented for entanglement purification \cite{EPP5}. The schematic diagram is depicted in Fig. 3. The pump pulse passes through the beta barium borate (BBO$_1$) and generates the state as \cite{EPP4}
\begin{eqnarray}\label{entanglementbyspdc}
|\Phi^+\rangle = |vac\rangle + \sqrt p |\phi^+\rangle + p |\phi^+\rangle^{ \otimes 2},
\end{eqnarray}
which is entangled in modes $a_1b_1$. Additionally, the pulse is reflected by a mirror to pass through the BBO$_1$ again to produce the entanglement in modes $a_2b_2$ like Eq. (\ref{entanglementbyspdc}). The bit-flip error occurs on the desired state with the probability of $1-F$ during the entanglement distribution, yielding a mixed state as
\begin{eqnarray}\label{mixedentanglementbyspdc}
\rho^{\prime}  = F| {{\Phi ^ + }}\rangle\langle {{\Phi ^ + }}| + (1 - F)| {{\Psi ^ + }}\rangle\langle {{\Psi ^ + }}|,
\end{eqnarray}
with
\begin{eqnarray}\label{bitflipentanglementbyspdc}
|{\Psi ^ + }\rangle  = |vac\rangle  + \sqrt p |{\psi ^ + }\rangle  + p|{\psi ^ + }{\rangle ^{ \otimes 2}}.
\end{eqnarray}
In this EPP, the ``four mode'' case is also used to herald a successful purification. Two photon state that both SPDC sources emit single-pair entanglement can be eliminated. Each SPDC source emits single-pair entanglement, which is the same as Ref. \cite{EPP3}. Thus, we only analyze the double-pair emissions generated from one SPDC source and the other one SPDC source emits the vacuum state in the following parts.

For the state $|\phi^+\rangle^{ \otimes 2}_{a_1b_1}$, it remains unchanged  with the probability of $Fp^2$ and becomes $|\psi^+\rangle^{ \otimes 2}_{a_1b_1}$ with the probability of $(1-F)p^2$. Hence, after passing through the PBSs, the state $|\phi^+\rangle^{ \otimes 2}_{a_1b_1}$ and $|\psi^+\rangle^{ \otimes 2}_{a_1b_1}$ will separately evolve to
\begin{eqnarray}\label{doublepair}
\!\!\!\!\frac{1}{2}{(|H\rangle _{{a_4}}}|H{\rangle _{{a_4}}}|H{\rangle _{{b_4}}}|H{\rangle _{{b_4}}} + |V{\rangle _{{a_3}}}|H{\rangle _{{a_4}}}|V{\rangle _{{b_3}}}|H{\rangle _{{b_4}}}\nonumber\\
 + |V{\rangle _{{a_3}}}|H{\rangle _{{a_4}}}|V{\rangle _{{b_3}}}|H{\rangle _{{b_4}}} + |V{\rangle _{{a_3}}}|V{\rangle _{{a_3}}}|V{\rangle _{{b_3}}}|V{\rangle _{{b_3}}}),
\end{eqnarray}
and
\begin{eqnarray}\label{doublepairbitflip}
\!\!\!\!\frac{1}{2}{(|H\rangle _{{a_4}}}|H{\rangle _{{a_4}}}|V{\rangle _{{b_3}}}|V{\rangle _{{b_3}}} + |V{\rangle _{{a_3}}}|H{\rangle _{{a_4}}}|V{\rangle _{{b_3}}}|H{\rangle _{{b_4}}}\nonumber\\
+|V{\rangle _{{a_3}}}|H{\rangle _{{a_4}}}|V{\rangle _{{b_3}}}|H{\rangle _{{b_4}}} + |V{\rangle _{{a_3}}}|V{\rangle _{{a_3}}}|H{\rangle _{{b_4}}}|H{\rangle _{{b_4}}}).
\end{eqnarray}
It is clear to observe that the item $|V{\rangle _{{a_3}}}|H{\rangle _{{a_4}}}|V{\rangle _{{b_3}}}|H{\rangle _{{b_4}}}$ makes each one of output modes contain one photon. Similarly, the photons of the states $|\phi^+\rangle^{ \otimes 2}_{a_2b_2}$ and $|\psi^+\rangle^{ \otimes 2}_{a_2b_2}$ after the PBSs may also result in the ``four mode'' case.  Moreover, one can make $|V{\rangle _{{a_3}}}|H{\rangle _{{a_4}}}|V{\rangle _{{b_3}}}|H{\rangle _{{b_4}}}$ and $|H{\rangle _{{a_3}}}|V{\rangle _{{a_4}}}|H{\rangle _{{b_3}}}|V{\rangle _{{b_4}}}$ in a coherent superposition provided that the amplitudes of these two ``four mode'' contributions simultaneously arrive at two PBSs. Finally, we use the basis $\{ | + \rangle ,| - \rangle \} $ to measure the photons in modes $a_4$ and $b_4$, yielding
\begin{eqnarray}\label{new}
{\rho _{{a_3}{b_3}}^{\prime}} = {F_2}|{\phi ^ + }{\rangle _{{a_3}{b_3}}}{\langle {\phi ^ + }| + (1 - {F_2})|{\psi ^ + }\rangle _{{a_3}{b_3}}}\langle {\psi ^ + }|.
\end{eqnarray}
In Ref. \cite{EPP5}, the initial fidelity of the mixed state is $3/4$. After one round of the successful purification, the fidelity can reach $13/14$. So far, we have reviewed the experiment of the EPP based on the PBSs. In addition to the purification experiments in optical system, some groups also carried out the EPP experiments in the atomic systems \cite{EPP51}, the solid-states \cite{EPP18Science} and the superconducting systems \cite{EPPPRL1}, respectively.
\begin{figure}
\begin{center}
\includegraphics[scale=1.9]{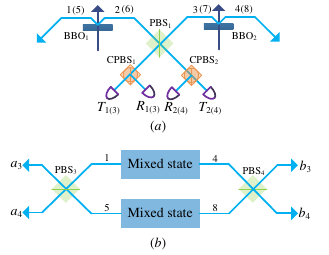}
\caption{The schematic diagram of the nested EPP for quantum repeater \cite{EPP16}. (a) The pump pulses pass through the BBO$_1$ and BBO$_2$ to generate the entanglement between the photons 1 and 2 (5 and 6), 3 and 4 (7 and 8). Subsequently, the entanglement swapping operations between the photons 2 and 3, 6 and 7 are respectively performed aimed to  eliminate the double-pair emissions from the SPDC sources provided that we pick out the case that coincidence detections on $T_1$ and $T_2$ or $R_1$ and $R_2$ ($T_3$ and $T_4$ or $R_3$ and $R_4$). (b) The photons 1 and 4, 5 and 8 are directed to the PBSs. Moreover, we select the ``four mode'' case that each of output modes $a_3$, $a_4$, $b_3$, and $b_4$ contain one photon. Finally, measurements in modes $a_4$ and $b_4$ with the basis $\{|+\rangle, |-\rangle\}$ make a projection into a high-quality entanglement. The circular PBS (CPBS) transmits the polarization $|+\rangle$ and reflects the photon in polarization $|-\rangle$. }
\end{center}
\end{figure}
\subsection{The nested EPP for quantum repeaters}
The EPP determines the efficiency of long-distance quantum communication. Consequently, Chen \emph{et al.} experimentally realized the nested entanglement purification for quantum repeaters in linear optics, in which the double-pair noise from the SPDC sources can be removed by entanglement swapping \cite{EPP16}. The four photon pairs such as 1 and 2, 3 and 4, 5 and 6, 7 and 8 have the same form as Eq. (\ref{entanglementbyspdc}), which are produced from the SPDC sources. After performing the entanglement swapping and selecting the case that the coincidence detections on $T_1$ and $T_2$ or $R_1$ and $R_2$ as depicted in Fig. 4(a), the state $|\Phi\rangle_{12}\otimes|\Phi\rangle_{34}$ evolves to
\begin{eqnarray}\label{11}
|{\phi _{14}}\rangle  &=& \frac{p}{2}{(|H\rangle _1}|H{\rangle _4} + |V{\rangle _1}|V{\rangle _4} + |H{\rangle _1}|H{\rangle _1}\nonumber\\
 &-&|V{\rangle _1}|V{\rangle _1} + |H{\rangle _4}|H{\rangle _4} - |V{\rangle _4}|V{\rangle _4}),
\end{eqnarray}
which automatically washes out spurious contributions resulted from the double-pair emissions. Similarly, if the coincidence detections on $T_3$ and $T_4$ or $R_3$ and $R_4$, indicating that the state $|\Phi\rangle_{56}\otimes|\Phi\rangle_{78}$ collapses to
\begin{eqnarray}\label{12}
|{\phi _{58}}\rangle  &=& \frac{p}{2}{(|H\rangle _5}|H{\rangle _8} + |V{\rangle _5}|V{\rangle _8} + |H{\rangle _5}|H{\rangle _5}\nonumber\\
 &-&|V{\rangle _5}|V{\rangle _5} + |H{\rangle _8}|H{\rangle _8} - |V{\rangle _8}|V{\rangle _8}).
\end{eqnarray}
The noise in quantum channels makes the state $|{\phi _{i,i+3}}\rangle$ ($i=1, 5$) become a mixed state written as
\begin{eqnarray}\label{14mixed}
{\rho _{i,i+3}} = F|{\phi _{{i,i+3}}}\rangle\langle {{\phi _{{i,i+3}}}} | + (1 - F)|{\psi _{{i,i+3}}}\rangle\langle {{\psi _{{i,i+3}}}}|,
\end{eqnarray}
with
\begin{eqnarray}\label{14psi}
|{\psi _{{i,i+3}}}\rangle  \!&=&\! \frac{p}{2}{(|HV\rangle _{i,i+3}} \!+\! |VH{\rangle _{i,i+3}} \!+\! |HH{\rangle _{i,i}}\nonumber\\
 \!&-&\!|VV{\rangle _{i,i}}\!+\! |VV{\rangle _{i+3,i+3}}\!-\!|HH{\rangle _{i+3,i+3}}).
\end{eqnarray}
Then, the same method as Refs. \cite{EPP3,EPP5} can be performed in a next step, i.e., the ``four mode'' case. Finally, the new mixed state with a higher fidelity can be obtained as the same form of Eq. (\ref{PBSnewmixed}).

\section{The EPP for multipartite systems}
This section provides an overview of the EPPs for multipartite systems. The first EPP for Greenberg-Horne-Zeilinger (GHZ) states was presented by Murao \emph{et al.} in 1998, named MMEPP \cite{EPP29}. In Refs. \cite{EPP30,EPP31,EPP32}, D\"ur \emph{et al.} presented a proposal to purify arbitrary two-colorable graph states including cluster states, GHZ states, and various error correcting codes. In Ref. \cite{EPP33}, Sheng \emph{et al.} used cross-Kerr nonlinearities to construct nondestructive quantum nondemolition (QND) instead of the CNOT gates and sophisticated single-photon detectors to purify the polluted GHZ states. For simplicity, we call it SMEPP. After that, Deng further proposed a high-efficient EPP named DMEPP for the multipartite entanglement including two steps \cite{EPP34}. The first step is the same as the SMEPP \cite{EPP33}. The second step reuses the discarding items in the SMEPP to generate entanglement. Recently, de Bone \emph{et al.} presented a proposal to create and distill the GHZ states from the noisy Bell states \cite{EPP36}. They employed a dynamic programming algorithm to minimize the consumed Bell states to perform creation and purification.

The remainder of this section will mainly review the MMEPP based on the CNOT gates \cite{EPP29} as well as the SMEPP using the QND \cite{EPP33} and the high-efficient DMEPP \cite{EPP34}.

\begin{figure}
\begin{center}
\includegraphics[scale=2.2]{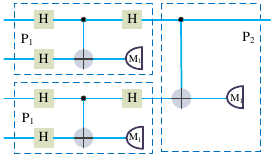}
\caption{The schematic diagram of the MMEPP \cite{EPP29}. This EPP can be divided into two steps such as P$_1$ correction for phase-flip errors and P$_2$ correction for bit-flip errors. Each round of this EPP requires four pairs of entangled states. The H denotes the Hadamard gate. }
\end{center}
\end{figure}

\subsection{The MMEPP based on the CNOT gates}
The schematic diagram of the MMEPP \cite{EPP29} is depicted in Fig. 5, which is made up of two parts that P$_1$ and P$_2$ respectively denote the corrections for phase-flip errors and bit-flip errors. Additionally, each party holds the same setup as Fig. 5. Now, let's take three-particle GHZ states as an example to briefly illustrate the principle of this MMEPP. Assume that the desired state is $|\phi^+_0\rangle_{abc}$, which is one of eight GHZ states as the form of
\begin{eqnarray}\label{GHZ}
&&|\phi _0^ \pm {\rangle _{abc}} = \frac{1}{{\sqrt 2 }}({| {000} \rangle _{abc}} \pm {| {111} \rangle _{abc}}),\nonumber\\
&&|\phi _1^ \pm {\rangle _{abc}} = \frac{1}{{\sqrt 2 }}({| {100} \rangle _{abc}} \pm {| {011} \rangle _{abc}}),\nonumber\\
&&|\phi _2^ \pm {\rangle _{abc}} = \frac{1}{{\sqrt 2 }}({| {010} \rangle _{abc}} \pm {| {101} \rangle _{abc}}),\nonumber\\
&&|\phi _3^ \pm {\rangle _{abc}} = \frac{1}{{\sqrt 2 }}({| {001} \rangle _{abc}} \pm {| {110} \rangle _{abc}}).
\end{eqnarray}
Due to the inherent noise in quantum channels, the bit-flip errors and phase-flip errors will occur on the initial state. For simplicity, we first consider that the phase-flip error occurs on the first particle with the probability of $1-F$, yielding
\begin{eqnarray}\label{GHZmixedstate1}
\rho^{{\prime}{\prime}}_p  = F|\phi _0^ + {\rangle _{abc}}\langle {\phi _0^ + }| + (1 - F)|\phi _0^ - {\rangle _{abc}}\langle {\phi _0^ - }|.
\end{eqnarray}
After performing the Hadamard operations, the states $|\phi _0^ + {\rangle _{abc}}$ and $|\phi _0^ - {\rangle _{abc}}$ separately become
\begin{eqnarray}\label{GHZevolvedstate1}
&&|\psi _0^ + {\rangle _{abc}} =\frac{1}{2}{(| {000} \rangle  + | {011} \rangle + | {101} \rangle + | {110} \rangle )},\nonumber\\
&&|\psi _0^ - {\rangle _{abc}}= \frac{1}{2}{(| {001} \rangle + | {010} \rangle + | {100} \rangle + | {111} \rangle )}.
\end{eqnarray}
In this way, the phase-flip error has been transformed to the bit-flip error as
\begin{eqnarray}\label{GHZmixedstate2}
\rho ^{{\prime}{\prime}}_b = F|\psi _0^ + {\rangle _{abc}}\langle {\psi _0^ + }| + (1 - F)|\psi _0^ - {\rangle _{abc}}\langle {\psi _0^ - }|.
\end{eqnarray}
After passing through the CNOT gates, we have
\begin{eqnarray}\label{GHZCNOT}
&&|\psi _0^ + {\rangle _{{a_1}{b_1}{c_1}}}|\psi _0^ + {\rangle _{{a_2}{b_2}{c_2}}} \to |\psi _0^ + {\rangle _{{a_1}{b_1}{c_1}}}|\psi _0^ + {\rangle _{{a_2}{b_2}{c_2}}},\nonumber\\
&&|\psi _0^ + {\rangle _{{a_1}{b_1}{c_1}}}|\psi _0^ - {\rangle _{{a_2}{b_2}{c_2}}} \to |\psi _0^ + {\rangle _{{a_1}{b_1}{c_1}}}|\psi _0^ - {\rangle _{{a_2}{b_2}{c_2}}},\nonumber\\
&&|\psi _0^ - {\rangle _{{a_1}{b_1}{c_1}}}|\psi _0^ + {\rangle _{{a_2}{b_2}{c_2}}} \to |\psi _0^ - {\rangle _{{a_1}{b_1}{c_1}}}|\psi _0^ - {\rangle _{{a_2}{b_2}{c_2}}},\nonumber\\
&&|\psi _0^ - {\rangle _{{a_1}{b_1}{c_1}}}|\psi _0^ - {\rangle _{{a_2}{b_2}{c_2}}} \to |\psi _0^ - {\rangle _{{a_1}{b_1}{c_1}}}|\psi _0^ + {\rangle _{{a_2}{b_2}{c_2}}}.
\end{eqnarray}
Then, by measuring the particles in the modes $a_2$, $b_2$, and $c_2$, we retain the source pair if the even number of $|1\rangle$ is obtained. Otherwise, we discard the source pair. Consequently, a new mixed state can be given by
\begin{eqnarray}\label{newGHZmixedstate}
\rho_{a_1b_1c_1}  \!=\! F_1|\psi _0^ + {\rangle _{a_1b_1c_1}}\langle {\psi _0^ + }| \!+\! (1 \!-\! F_1)|\psi _0^ - {\rangle _{a_1b_1c_1}}\langle {\psi _0^ - }|.
\end{eqnarray}
By adding the Hadamard operations on the particles in modes $a_1b_1c_1$, we can respectively transform the state $|\psi _0^ + {\rangle _{a_1b_1c_1}}$ and $|\psi _0^ - {\rangle _{a_1b_1c_1}}$ to $|\phi _0^ + {\rangle _{a_1b_1c_1}}$ and $|\phi _0^ - {\rangle _{a_1b_1c_1}}$.

For the bit-flip error, one can use P$_2$ to directly purify it. Suppose that the mixed state is
\begin{eqnarray}\label{bitflipGHZmixedstate}
\rho^{{\prime}{\prime}}  = F|\phi _0^ + {\rangle _{abc}}\langle {\phi _0^ + } | + (1 - F)|\phi _1^ + {\rangle _{abc}}\langle {\phi _1^ + } | .
\end{eqnarray}
After performing the CNOT operations, one can obtain
\begin{eqnarray}\label{GHZCNOT1}
&&|\phi _0^ + {\rangle _{{a_1}{b_1}{c_1}}}|\phi _0^ + {\rangle _{{a_2}{b_2}{c_2}}} \to |\phi _0^ + {\rangle _{{a_1}{b_1}{c_1}}}|\phi _0^ + {\rangle _{{a_2}{b_2}{c_2}}},\nonumber\\
&&|\phi _0^ + {\rangle _{{a_1}{b_1}{c_1}}}|\phi _1^ + {\rangle _{{a_2}{b_2}{c_2}}} \to |\phi _0^ + {\rangle _{{a_1}{b_1}{c_1}}}|\phi _1^ + {\rangle _{{a_2}{b_2}{c_2}}},\nonumber\\
&&|\phi _1^ + {\rangle _{{a_1}{b_1}{c_1}}}|\phi _0^ + {\rangle _{{a_2}{b_2}{c_2}}} \to |\phi _1^ + {\rangle _{{a_1}{b_1}{c_1}}}|\phi _1^ + {\rangle _{{a_2}{b_2}{c_2}}},\nonumber\\
&&|\phi _1^ + {\rangle _{{a_1}{b_1}{c_1}}}|\phi _1^ + {\rangle _{{a_2}{b_2}{c_2}}} \to |\phi _1^ + {\rangle _{{a_1}{b_1}{c_1}}}|\phi _0^ + {\rangle _{{a_2}{b_2}{c_2}}}.
\end{eqnarray}
If the same measurement result is obtained, we remain the source pair. Otherwise, we discard the source pair. By far, we have briefly introduced the MMEPP \cite{EPP29} for the three-particle GHZ states and the similar analysis can be extended to the arbitrary multipartite GHZ states.
\begin{figure}
\begin{center}
\includegraphics[scale=2.2]{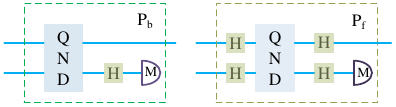}
\caption{The schematic diagram of the SMEPP \cite{EPP33}. P$_b$ is the setup for correcting bit-flip errors and P$_f$ is the setup for correcting phase-flip errors. The QND is shown in Fig. 7.}
\end{center}
\end{figure}

\subsection{The SMEPP based on the QND}
The principle of the SMEPP \cite{EPP33} based on the QND is depicted in Fig. 6. The QND as shown in Fig. 7 not only plays the role of the CNOT gate but also single-photon measurement. To be specific, if two photons are in $|HH\rangle$ or $|VV\rangle$, the phase shift carried by the coherent state is $\theta$. If they are in $|HV\rangle$ or $|VH\rangle$, the phase shift carried by the coherent state is $2\theta$ or 0, respectively.

\begin{figure}
\begin{center}
\includegraphics[scale=2.2]{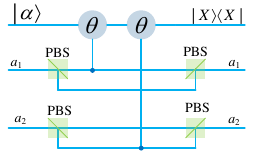}
\caption{The schematic diagram of QND \cite{EPP33} in Fig. 6. This QND can deterministically distinguish $|HH\rangle$ and $|VV\rangle$ from $|HV\rangle$ and $|VH\rangle$.  }
\end{center}
\end{figure}

We first illustrate the principle of correction for bit-flip errors such as P$_b$. The mixed state is given by
\begin{eqnarray}\label{PhotonGHZMixed}
{\rho^b _{{a_1}{b_1}{c_1}}} = F|\phi _4^ + {\rangle _{{a_1}{b_1}{c_1}}}{\langle \phi _4^ + | + (1 - F)|\phi _5^ + \rangle _{{a_1}{b_1}{c_1}}}\langle \phi _5^ + |,
\end{eqnarray}
where
\begin{eqnarray}\label{PhotonGHZ}
&&|\phi _4^ \pm {\rangle _{abc}} = \frac{1}{{\sqrt 2 }}{(|HHH\rangle _{abc}} \pm |VVV{\rangle _{abc}}), \nonumber \\
&&|\phi _5^ \pm {\rangle _{abc}} = \frac{1}{{\sqrt 2 }}{(|VHH\rangle _{abc}} \pm |HVV{\rangle _{abc}}), \nonumber \\
&&|\phi _6^ \pm {\rangle _{abc}} = \frac{1}{{\sqrt 2 }}{(|HVH\rangle _{abc}} \pm |VHV{\rangle _{abc}}), \nonumber \\
&&|\phi _7^ \pm {\rangle _{abc}} = \frac{1}{{\sqrt 2 }}{(|VVH\rangle _{abc}} \pm |HHV{\rangle _{abc}}).
\end{eqnarray}
After passing through the QND, three parties pick out the cases that the phase shifts of their coherent states are $\theta$, this makes the states $|\phi _4^ + {\rangle _{a_1b_1c_1}}\otimes|\phi _4^ + {\rangle _{a_2b_2c_2}}$ and $|\phi _5^ + {\rangle _{a_1b_1c_1}}\otimes|\phi _5^ + {\rangle _{a_2b_2c_2}}$ project into
\begin{eqnarray}\label{evolveQND1}
|{\phi _4}\rangle  \!=\! \frac{1}{{\sqrt 2 }}(| {HHHHHH}\rangle  \!\!+\!\! | {VVVVVV} \rangle )_{{a_1}{b_1}{c_1}{a_2}{b_2}{c_2}},
\end{eqnarray}
and
\begin{eqnarray}\label{evolveQND2}
|{\phi _5}\rangle  \!=\! \frac{1}{{\sqrt 2 }}(| {VHHVHH}\rangle  \!\!+\!\! | {HVVHVV} \rangle )_{{a_1}{b_1}{c_1}{a_2}{b_2}{c_2}},
\end{eqnarray}
with the probability of $\frac{{{F^2}}}{2}$ and $\frac{{{(1-F)^2}}}{2}$, respectively. Moreover, the cross combinations $|\phi _4^ + {\rangle _{a_1b_1c_1}}\otimes|\phi _5^ + {\rangle _{a_2b_2c_2}}$ and $|\phi _5^ + {\rangle _{a_1b_1c_1}}\otimes|\phi _4^ + {\rangle _{a_2b_2c_2}}$ can be discarded according to the results of the $X$ homodyne measurements. After that, adding the Hadamard operations on the photons in modes $a_2b_2c_2$ followed by measuring these photons with the Z-basis, one can get a new mixed state with the fidelity of $F_1$ provided that the number of $|V\rangle$ is even. While if it is odd, an additional phase-flip operation should be performed on one of the photons. If the phase-flip error occurs, the mixed state is
\begin{eqnarray}\label{PhotonGHZPhase}
{\rho^p _{abc}} = F|\phi _4^ + {\rangle _{abc}}{\langle \phi _4^ + | + (1 - F)|\phi _4^ - \rangle _{abc}}\langle \phi _4^ - |
\end{eqnarray}
After performing the Hadamard operations, ${\rho^p _{abc}}$ becomes

\begin{eqnarray}\label{PhotonGHZPhase1}
&&|\psi _4^ + {\rangle _{abc}} = \frac{1}{2}(|HHH\rangle  + |HVV\rangle  + |VHV\rangle  + |VVH\rangle ), \nonumber \\
&&|\psi _4^ - {\rangle _{abc}} = \frac{1}{2}(|HHV\rangle  + |HVH\rangle  + |VHH\rangle  + |VVV\rangle ). \nonumber \\
\end{eqnarray}
Then the same method as P$_b$ can be done in a next step. The success probability of the SMEPP based on the QND is $\frac{{{F^2} + {{(1 - F)}^2}}}{2}$. However, if we make $\theta  = \pi$, the $2\theta$ and 0 can not be distinguished. Hence, the discarding components of $|\phi _4^ + {\rangle _{a_1b_1c_1}}\otimes|\phi _4^ + {\rangle _{a_2b_2c_2}}$ and $|\phi _5^ + {\rangle _{a_1b_1c_1}}\otimes|\phi _5^ + {\rangle _{a_2b_2c_2}}$ can contribute to the SMEPP. As a result, the efficiency of the SMEPP can be doubled provided that $\theta  = \pi$.

\subsection{The high-efficient DMEPP}
In 2011, Deng proposed a high-efficient DMEPP scheme for the multipartite entanglement including two steps \cite{EPP34}. The first step is the same as the SMEPP in Ref. \cite{EPP33}. The second step reuses the discarding items in the SMEPP to produce entanglement. For example, the DMEPP considered a general mixed state given by
\begin{eqnarray}\label{arbitrarymixed}
\rho^{{\prime}{\prime}{\prime}}  &=& F|\phi _4^ + {\rangle _{abc}}{\langle \phi _4^ + | + {F_0}|\phi _5^ + \rangle _{abc}}\langle \phi _5^ + |\nonumber\\
&+&{F_2}|\phi _6^ + {\rangle _{abc}}{\langle \phi _6^ + | + {F_3}|\phi _7^ + \rangle _{abc}}\langle \phi _7^ + |.
\end{eqnarray}
After performing the first step of the DMEPP, a new mixed state can be obtained as
\begin{eqnarray}\label{newarbitrarymixed}
\rho^{{\prime}{\prime}{\prime}}_1  &=&{F^\prime }|\phi _4^ + {\rangle _{abc}}{\langle \phi _4^ + | + {F^\prime_0 }|\phi _5^ + \rangle _{abc}}\langle \phi _5^ + |\nonumber\\
&+&{F^\prime_2 }|\phi _6^ + {\rangle _{abc}}{\langle \phi _6^ + | + {F^\prime_3 }|\phi _7^ + \rangle _{abc}}\langle \phi _7^ + |,
\end{eqnarray}
with ${F^\prime } = \frac{{{F^2}}}{N}$, ${F^\prime_0 } = \frac{{{F^2_0}}}{N}$, ${F^\prime_2 } = \frac{{{F^2_2}}}{N}$, and ${F^\prime_3 } = \frac{{{F^2_3}}}{N}$, where $N = {F^2} + F_0^2 + F_2^2 + F_3^2$.

\begin{figure}
\begin{center}
\includegraphics[scale=2.2]{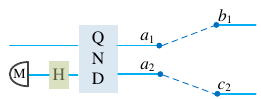}
\caption{The schematic diagram of the second step of the DMEPP \cite{EPP34}.}
\end{center}
\end{figure}

In the second step, the discarding items can be reused to generate the GHZ states. Here we take $|\phi _4^ + {\rangle _{abc}}|\phi _6^ + {\rangle _{abc}}$, $|\phi _6^ + {\rangle _{abc}}|\phi _4^ + {\rangle _{abc}}$, $|\phi _5^ + {\rangle _{abc}}|\phi _7^ + {\rangle _{abc}}$, and $|\phi _7^ + {\rangle _{abc}}|\phi _5^ + {\rangle _{abc}}$ as examples to illustrate the principle of the second step. If the measurement outcomes of Alice, Bob, and Charlie are even mode, odd mode, and even mode, $|\phi _4^ + {\rangle _{abc}}|\phi _6^ + {\rangle _{abc}}$, $|\phi _6^ + {\rangle _{abc}}|\phi _4^ + {\rangle _{abc}}$, $|\phi _5^ + {\rangle _{abc}}|\phi _7^ + {\rangle _{abc}}$, and $|\phi _7^ + {\rangle _{abc}}|\phi _5^ + {\rangle _{abc}}$ separately evolve to
\begin{eqnarray}\label{xi1}
{\xi _1} = \frac{1}{{\sqrt 2 }}({| {HHH}\rangle _{{a_1}{b_1}{c_1}}}{| {HVH} \rangle _{{a_2}{b_2}{c_2}}}\nonumber\\
 + {| {VVV} \rangle _{{a_1}{b_1}{c_1}}}{| {VHV} \rangle _{{a_2}{b_2}{c_2}}}),\nonumber\\
{\xi _2} = \frac{1}{{\sqrt 2 }}({| {HVH}\rangle _{{a_1}{b_1}{c_1}}}{| {HHH} \rangle _{{a_2}{b_2}{c_2}}}\nonumber\\
 + {| {VHV} \rangle _{{a_1}{b_1}{c_1}}}{| {VVV} \rangle _{{a_2}{b_2}{c_2}}}),\nonumber\\
{\zeta _1} = \frac{1}{{\sqrt 2 }}({| {VHH}\rangle _{{a_1}{b_1}{c_1}}}{| {VVH}\rangle _{{a_2}{b_2}{c_2}}}\nonumber\\
 + {| {HVV}\rangle _{{a_1}{b_1}{c_1}}}{| {HHV}\rangle _{{a_2}{b_2}{c_2}}}),\nonumber\\
 {\zeta _2} = \frac{1}{{\sqrt 2 }}({| {HHV}\rangle _{{a_1}{b_1}{c_1}}}{| {HVV}\rangle _{{a_2}{b_2}{c_2}}}\nonumber\\
 + {| {VVH}\rangle _{{a_1}{b_1}{c_1}}}{| {VHH}\rangle _{{a_2}{b_2}{c_2}}}),
\end{eqnarray}
with the probability of $\frac{{F{F_2}}}{2}$, $\frac{{F{F_2}}}{2}$, $\frac{{{F_0}{F_3}}}{2}$, and $\frac{{{F_0}{F_3}}}{2}$. Similarly, if the measurement outcomes are odd mode, even mode, and odd mode, they collapse to
\begin{eqnarray}\label{xi2}
{\xi _3} = \frac{1}{{\sqrt 2 }}({| {HHH} \rangle _{{a_1}{b_1}{c_1}}}{| {VHV} \rangle _{{a_2}{b_2}{c_2}}}\nonumber\\
 + {| {VVV} \rangle _{{a_1}{b_1}{c_1}}}{| {HVH} \rangle _{{a_2}{b_2}{c_2}}}),\nonumber\\
{\xi _4} = \frac{1}{{\sqrt 2 }}({| {HVH} \rangle _{{a_1}{b_1}{c_1}}}{| {VVV} \rangle _{{a_2}{b_2}{c_2}}}\nonumber\\
 + {| {VHV} \rangle _{{a_1}{b_1}{c_1}}}{| {HHH} \rangle _{{a_2}{b_2}{c_2}}}),\nonumber\\
{\zeta _3} = \frac{1}{{\sqrt 2 }}({| {VHH}\rangle _{{a_1}{b_1}{c_1}}}{| {HHV} \rangle _{{a_2}{b_2}{c_2}}}\nonumber\\
 + {| {HVV} \rangle _{{a_1}{b_1}{c_1}}}{| {VVH} \rangle _{{a_2}{b_2}{c_2}}}),\nonumber\\
{\zeta _4} = \frac{1}{{\sqrt 2 }}({| {HHV}\rangle _{{a_1}{b_1}{c_1}}}{| {VHH} \rangle _{{a_2}{b_2}{c_2}}}\nonumber\\
 + {| {VVH} \rangle _{{a_1}{b_1}{c_1}}}{| {HVV} \rangle _{{a_2}{b_2}{c_2}}}),
\end{eqnarray}
with the probability of $\frac{{F{F_2}}}{2}$, $\frac{{F{F_2}}}{2}$, $\frac{{{F_0}{F_3}}}{2}$, and $\frac{{{F_0}{F_3}}}{2}$. Subsequently, we measure the photons in modes $b_1$, $a_2$, $b_2$, and $c_2$ with the basis $\{ {| + \rangle ,| - \rangle }\}$. If the number of measurement outcomes $|-\rangle$ is even, we obtain a new mixed state. If the number of measurement outcomes $|-\rangle$ is odd, it requires an additional operation $\sigma_z$ to be performed on one of the photons. Hence, we can obtain
\begin{eqnarray}\label{new1}
\rho_{a_1c_1}  = 2FF_2|{\phi ^ + }{\rangle _{{a_1}{c_1}}}{\langle {\phi ^ + }| + 2F_0F_3|{\psi ^ + }\rangle _{{a_1}{c_1}}}\langle {\psi ^ + }|.
\end{eqnarray}
If $F > {F_0} = {F_2} = {F_3}$, we have $\frac{F}{{F + {F_0}}} > F$.
Similarly, the other discarding components such as
\begin{eqnarray}\label{new1}
\rho_{a_1b_1}  \!&=&\! 2FF_3|{\phi ^ + }{\rangle _{{a_1}{b_1}}}{\langle {\phi ^ + }| \!+\! 2F_0F_2|{\psi ^ + }\rangle _{{a_1}{b_1}}}\langle {\psi ^ + }|,\nonumber\\
\rho_{b_1c_1}  \!&=&\! 2FF_0|{\phi ^ + }{\rangle _{{b_1}{c_1}}}{\langle {\phi ^ + }| \!+\! 2F_2F_3|{\psi ^ + }\rangle _{{b_1}{c_1}}}\langle {\psi ^ + }|,
\end{eqnarray}
can be reused to produce entanglement. Interestingly, the high-fidelity of the GHZ states compared to the initial one can be produced using these Bell states. For instance, the system $\rho_{a_1b_1}\otimes\rho_{a_2c_2}$ is made up of four pure states such as $|{\phi ^ + }{\rangle _{{a_1}{b_1}}}|{\phi ^ + }{\rangle _{{a_2}{c_2}}}$, $|{\phi ^ + }{\rangle _{{a_1}{b_1}}}|{\psi ^ + }{\rangle _{{a_2}{c_2}}}$, $|{\psi ^ + }{\rangle _{{a_1}{b_1}}}|{\phi ^ + }{\rangle _{{a_2}{c_2}}}$, and $|{\psi ^ + }{\rangle _{{a_1}{b_1}}}|{\psi ^ + }{\rangle _{{a_2}{c_2}}}$ with the probability of $\frac{{{F^2}}}{{{{(F + {F_0})}^2}}}$, $\frac{{{FF_0}}}{{{{(F + {F_0})}^2}}}$, $\frac{{{FF_0}}}{{{{(F + {F_0})}^2}}}$, and $\frac{{{F_0^2}}}{{{{(F + {F_0})}^2}}}$, respectively. After passing through the setup shown in Fig. 8, if Alice' s measurement outcome is in even mode, the four states evolve to
\begin{eqnarray}\label{Belleven}
|{\Omega _1}\rangle  = \frac{1}{{\sqrt 2 }}{(|HHHH\rangle _{{a_1}{b_1}{a_2}{c_2}}} + |VVVV{\rangle _{{a_1}{b_1}{a_2}{c_2}}}),\nonumber\\
|{\Omega _2}\rangle  = \frac{1}{{\sqrt 2 }}{(|HHHV\rangle _{{a_1}{b_1}{a_2}{c_2}}} + |VVVH{\rangle _{{a_1}{b_1}{a_2}{c_2}}}),\nonumber\\
|{\Omega _3}\rangle  = \frac{1}{{\sqrt 2 }}{(|HVHH\rangle _{{a_1}{b_1}{a_2}{c_2}}} + |VHVV{\rangle _{{a_1}{b_1}{a_2}{c_2}}}),\nonumber\\
|{\Omega _4}\rangle  = \frac{1}{{\sqrt 2 }}{(|HVHV\rangle _{{a_1}{b_1}{a_2}{c_2}}} + |VHVH{\rangle _{{a_1}{b_1}{a_2}{c_2}}}),
\end{eqnarray}
with the probability of $\frac{{{F^2}}}{{{{2(F + {F_0})}^2}}}$, $\frac{{{FF_0}}}{{{{2(F + {F_0})}^2}}}$, $\frac{{{FF_0}}}{{{{2(F + {F_0})}^2}}}$, and $\frac{{{F_0^2}}}{{{{2(F + {F_0})}^2}}}$. Otherwise, the four states evolve to
\begin{eqnarray}\label{Bellodd}
|{\kappa _1}\rangle  = \frac{1}{{\sqrt 2 }}{(|HHVV\rangle _{{a_1}{b_1}{a_2}{c_2}}} + |VVHH{\rangle _{{a_1}{b_1}{a_2}{c_2}}}),\nonumber\\
|{\kappa _2}\rangle  = \frac{1}{{\sqrt 2 }}{(|HHVH\rangle _{{a_1}{b_1}{a_2}{c_2}}} + |VVHV{\rangle _{{a_1}{b_1}{a_2}{c_2}}}),\nonumber\\
|{\kappa _3}\rangle  = \frac{1}{{\sqrt 2 }}{(|HVVV\rangle _{{a_1}{b_1}{a_2}{c_2}}} + |VHHH{\rangle _{{a_1}{b_1}{a_2}{c_2}}}),\nonumber\\
|{\kappa _4}\rangle  = \frac{1}{{\sqrt 2 }}{(|HVVH\rangle _{{a_1}{b_1}{a_2}{c_2}}} + |VHHV{\rangle _{{a_1}{b_1}{a_2}{c_2}}}).
\end{eqnarray}
with the probability of $\frac{{{F^2}}}{{{{2(F + {F_0})}^2}}}$, $\frac{{{FF_0}}}{{{{2(F + {F_0})}^2}}}$, $\frac{{{FF_0}}}{{{{2(F + {F_0})}^2}}}$, and $\frac{{{F_0^2}}}{{{{2(F + {F_0})}^2}}}$. Then, after performing the Hadamard operation on the photon in mode $a_2$ and measuring the photon with the Z-basis, one can obtain
\begin{eqnarray}\label{new2}
{\rho _{{a_1}{b_1}{c_2}}^{\prime}} = {F^{{\prime}{\prime}} }|{\phi_{4} ^ + }{\rangle _{{a_1}{b_1}{c_2}}}{\langle {\phi_{4} ^ + }| + F_0^{{\prime}{\prime}} |\phi _5^ + \rangle _{{a_1}{b_1}{c_2}}}\langle \phi _5^ + |\nonumber\\
 + F_2^{{\prime}{\prime}} |\phi _6^ + {\rangle _{{a_1}{b_1}{c_2}}}{\langle \phi _6^ + | + F_2^{{\prime}{\prime}} |\phi _7^ + \rangle _{{a_1}{b_1}{c_2}}}\langle \phi _7^ + |,
\end{eqnarray}
where ${F^{{\prime}{\prime}} } = \frac{{{F^2}}}{{{{(F + {F_0})}^2}}}$, $F_0^{{\prime}{\prime}}  = \frac{{F_0^2}}{{{{(F + {F_0})}^2}}}$, and $F_2^{{\prime}{\prime}}  = \frac{{F{F_0}}}{{{{(F + {F_0})}^2}}}$. Here the fidelity of the resultant state from two pairs of Bell states ${F^{{\prime}{\prime}} }>F$ if $F >0.25$. We consider that $F_0=F_2=F_3=\frac{{1 - F}}{3}$, the efficiency of this DMEPP and the conventional EPP for multiparty systems \cite{EPP29,EPP33} can be given by $\frac{{1 - 2F + 4{F^2}}}{3}$ and $\frac{{2 - F + 2{F^2}}}{3}$, respectively. The fidelity of the conventional MEPP \cite{EPP29,EPP33} and the DMEPP can be formulated as $\frac{{3{F^2}}}{{1 - 2F + 4{F^2}}}$ and $\frac{{3{F^2}(4 + 7F - 2{F^2})}}{{{{(1 + 2F)}^2}(2 - F + 2{F^2})}}$. Fig. 9 plots the curves of efficiency and fidelity versus the initial fidelity. It is clear to observe that the efficiency of the   DMEPP \cite{EPP34} outperforms the conventional MEPP \cite{EPP29,EPP33} at the cost of fidelity slightly.

\begin{figure}
\begin{center}
\includegraphics[scale=0.5]{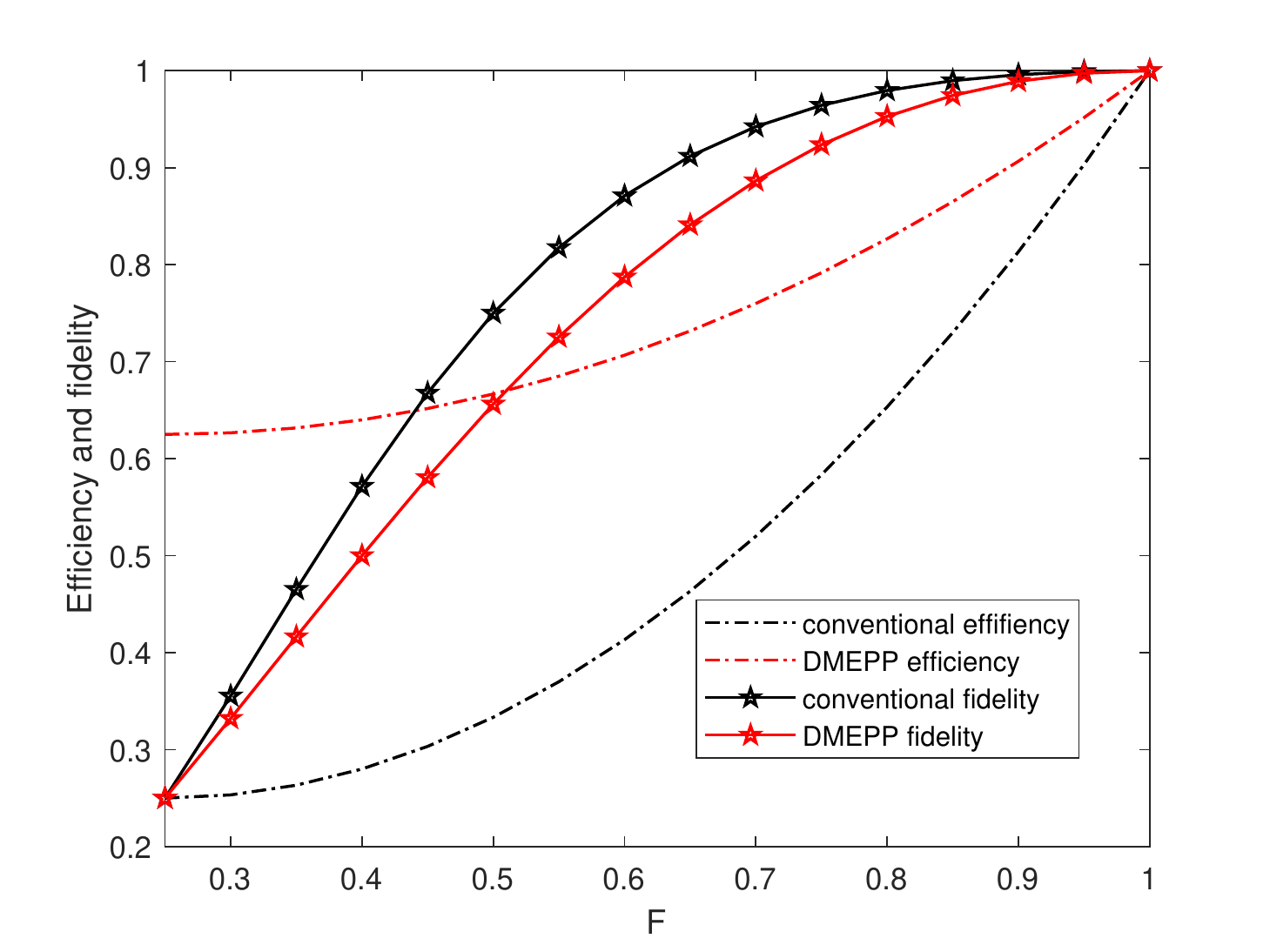}
\caption{The fidelities and efficiency of the DMEPP \cite{EPP34} and the conventional EPP for multiparty systems \cite{EPP29,EPP33} versus the initial fidelity. The red and black dotted lines (star lines) represent the efficiency (fidelities) of the DMEPP and the conventional EPP for multiparty systems.}
\end{center}
\end{figure}

\section{The Hyperentanglement EPP}
Hyperentanglement, the simultaneous entanglement
in more than one DOFs,  is widely explored in many quantum information protocols \cite{Hyperentanglementreview1}. For instance, Schuck \emph{et al.} reported an experiment that complete polarization Bell-state analysis (BSA) can be achieved with the help of   time-bin entanglement \cite{Hyperentanglement2}. In 2008, Barreiro \emph{et al.} employed hyperentanglement in polarization-orbital angular momentum to beat the channel capacity limit for superdense coding \cite{Hyperentanglement3}. In 2010, Sheng \emph{et al.} first used cross-Kerr nonlinearities to unambiguously distinguish the polarization-spatial mode hyperentanglement \cite{Hyperentanglement4}. In the later, Ren \emph{et al.} extended the results in Ref. \cite{Hyperentanglement4} to the giant nonlinear optics in quantum dot-cavity systems \cite{Hyperentanglement5} and Liu \emph{et al.} resorted the nitrogen-vacancy in micro-toroidal resonators to deterministically realize a hyperentanglement BSA \cite{Hyperentanglement6}.

Additionally, hyperentanglement can also be used to perform the entanglement purification. In 2002, Simon and Pan first employed the spatial mode entanglement to purify the polarization entanglement, named SPEPP \cite{EPP4}. In 2008, Sheng \emph{et al.} employed the QND to achieve an efficient EPP \cite{EPP6}. In 2010, Sheng and Deng proposed the concept of the deterministic entanglement purification with hyperentanglement \cite{EPP7}. In the same year,  one-step deterministic polarization EPPs resorting to the spatial entanglement in linear optics were proposed \cite{EPP8,Hyperentanglement8}. For multipartite systems, Deng \cite{Hyperentanglement81} and Sheng \emph{et al.} \cite{Hyperentanglement9} extended the one-step deterministic EPP to correct multipartite polarization entanglement consuming spatial entanglement.  In 2014, Sheng \emph{et al.} proposed another deterministic EPP in which the robust time-bin entanglement is regarded as resources to purify polarization entanglement \cite{Hyperentanglement11}. In Ref. \cite{Hyperentanglement12}, Ren \emph{et al.} employed the property of giant optical circular birefringence of a double-sided quantum dot (QD) cavity system to construct parity-check gate and quantum state joining method (QSJM) to realize a two-step hyperentanglement EPP (HEPP), which increases the fidelity of polarization-spatial hyperentanglement. In 2015, a new HEPP for overcoming the photon loss and decoherence was proposed with the assistance of the QND parity-checking measurement and the heralded two-qubit amplification \cite{Hyperentanglement13}. Shortly after that, they further proposed an efficient HEPP with imperfect spatial entanglement resources and relaxed the requirements for the HEPP using the high-dimensional mode-check
measurement \cite{WangOE2}. In 2016, Wang \emph{et al.} proposed a HEPP for three DOFs of two photons \cite{Hyperentanglement14}.
In 2021, a high-efficient EPP was realized which uses the noisy spatial entanglement to purify the polarization entanglement \cite{EPP26}. In this experiment, the purified entanglement can be further applied to the entanglement-based QKD and the secrecy key rate can be improved from 0 to 0.332. Moreover, the efficiency of this EPP is several orders of magnitude higher compared to the conventional EPPs using two noisy copies. Similarly, the noisy time-bin entanglement was exploited to purify polarization entanglement in experiment \cite{EPPadd1}. Subsequently, Zhou \emph{et al.} extended this EPP to purify the multi-particle entanglement \cite{Hyperentanglement16}. In \cite{Hyperentanglement17}, they further employed spatial entanglement and time-bin entanglement to purify the bit-flip errors and phase-flip errors of the polarization entanglement, respectively.

In the following parts, we will mainly review the SPEPP \cite{EPP4}, the high-efficient EPP with the QND \cite{EPP6}, the one-step deterministic EPP \cite{EPP8} and the single-copy high-efficient EPP \cite{EPP26} as well as the HEPP \cite{Hyperentanglement12}.

\subsection{The SPEPP in linear optics}
This subsection will briefly review the SPEPP which purifies polarization entanglement at the cost of spatial entanglement \cite{EPP4}. The schematic diagram of the SPEPP is shown in Fig. 3. It picks out the cases that both the photons come from both the upper modes or the lower modes.  As pointed out in Ref. \cite{EPP4}, the Hamiltonian of the SPDC source can be given by
\begin{eqnarray}\label{SPDC}
{H_{\text{spdc}}} &=& \delta [(a_{1H}^\dag b_{1H}^\dag  + a_{1V}^\dag b_{1V}^\dag )\nonumber\\
&+& \eta {e^{i\omega }}(a_{2H}^\dag b_{2H}^\dag  + a_{2V}^\dag b_{2V}^\dag )] + H.c..
\end{eqnarray}
For simplicity, let $\eta=1$ and $\omega=0$. Hence, the single-pair entanglement generated from the SPDC source shown in Fig. 3 can be written as $(a_{1H}^\dag b_{1H}^\dag  + a_{1V}^\dag b_{1V}^\dag  + a_{2H}^\dag b_{2H}^\dag  + a_{2V}^\dag b_{2V}^\dag )|0\rangle$. Actually, it is a hyperentangled state in polarization and spatial mode. The SPEPP is a success provided that the photons emit from the same output mode, that is, the upper mode or the lower mode.

As shown in Fig. 3, if the initial state is not polluted, it evolves to $(a_{3H}^\dag b_{3H}^\dag  + a_{3V}^\dag b_{3V}^\dag  + a_{4H}^\dag b_{4H}^\dag  + a_{4V}^\dag b_{4V}^\dag )|0\rangle$ after the PBSs, which indicates that the photons of Alice and Bob either come from the upper modes or the lower modes. If the desired state suffers from noise, i.e., bit-flip errors, it becomes $(a_{1H}^\dag b_{1V}^\dag  + a_{1V}^\dag b_{1H}^\dag  + a_{2H}^\dag b_{2V}^\dag  + a_{2V}^\dag b_{2H}^\dag )|0\rangle$. Thus, it collapses to $(a_{3H}^\dag b_{4V}^\dag  + a_{3V}^\dag b_{4H}^\dag  + a_{4H}^\dag b_{3V}^\dag  + a_{4V}^\dag b_{3H}^\dag )|0\rangle$ by the action of the PBSs. Obviously, one of two photons in the upper mode and the other one in the lower mode. As a result, bit-flip errors can be perfectly corrected according to the selection rule. However, phase-flip errors cannot be directly corrected because the spatial-mode entanglement has been consumed after purification for bit-flip errors. Under this circumstance, phase-flip errors should be transformed to bit-flip errors and the same method can be employed in a next step.

For the four-photon cases, that is each SPDC source generates one photon pair or one SPDC source generates two photon pairs while the other one produces vacuum state, i.e., ${(a_{1H}^\dag b_{1H}^\dag  + a_{1V}^\dag b_{1V}^\dag  + a_{2H}^\dag b_{2H}^\dag  + a_{2V}^\dag b_{2V}^\dag )^2}|0\rangle$, each output mode precisely contains one photon indicating a successful purification, which is the ``four mode'' case like the EPPs in Refs. \cite{EPP3,EPP5}.

\subsection{The polarization EPP based on the QND with the SPDC source }
\begin{figure}
\begin{center}
\includegraphics[scale=2.1]{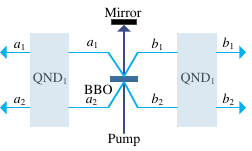}
\caption{The EPP based on the QND$_1$ with the SPDC sources \cite{EPP6}. The QND$_1$ plays the role of the CNOT gates and the photon-number detectors. The principle of the QND$_1$ is depicted in Fig. 11.}
\end{center}
\end{figure}
 In 2008, Sheng \emph{et al.}  presented an efficient EPP, which employs the QND to play the role of the CNOT gate and the photon-number detector \cite{EPP6}. The schematic diagram of this EPP is shown in Fig. 10. The SPDC source is similar to Fig. 3, which probabilistically generates two-photon state with the probability of $p$ and four-photon state with the probability of $p^2$.

\begin{figure}[!h]
\begin{center}
\includegraphics[scale=2.1]{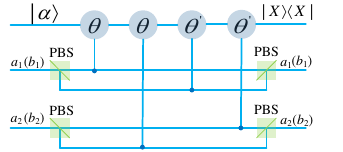}
\caption{The schematic diagram of the QND$_1$ \cite{EPP6}. It can distinguish the state $|HH\rangle$ and $|VV\rangle$ from $|HV\rangle$ and $|VH\rangle$. For example, if the phase shifts of coherent states for Alice and Bob are the same, the state is in $|HH\rangle$ or $|VV\rangle$. If Alice and Bob get different phase shifts, the state is in either $|HV\rangle$ or $|VH\rangle$. }
\end{center}
\end{figure}

For the two-photon state without encountering with bit-flip errors and phase-flip errors, the photons of two parties are in the same polarization. The same phase-shift can be obtained after the $X$ homodyne measurement shown in Fig. 11. After the action of the coupler shown in Fig. 12, they will either appear at the upper modes $a_1b_1$ or the lower modes $a_2b_2$. If the bit-flip error occurs on the desired state, $(a_{1H}^\dag b_{1H}^\dag  + a_{1V}^\dag b_{1V}^\dag  + a_{2H}^\dag b_{2H}^\dag  + a_{2V}^\dag b_{2V}^\dag )|0\rangle$ becomes $(a_{1V}^\dag b_{1H}^\dag  + a_{1H}^\dag b_{1V}^\dag  + a_{2V}^\dag b_{2H}^\dag  + a_{2H}^\dag b_{2V}^\dag )|0\rangle$. Clearly, two photons are in the different polarizations. As a result, the different phase shifts will be carried by Alice and Bob, indicating the presence of bit-flip errors. In this case, the bit-flip operation ${\sigma _x} = |H\rangle\langle V| + |V\rangle\langle H|$ is required to be performed on one particle to recover the initial one. Hence, the bit-flip error of the two-photon state can be perfectly corrected.

\begin{figure}
\begin{center}
\includegraphics[scale=1.8]{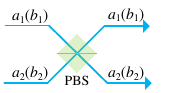}
\caption{The schematic diagram of the principle of a coupler \cite{EPP6}.}
\end{center}
\end{figure}

\begin{figure}
\begin{center}
\includegraphics[scale=2.1]{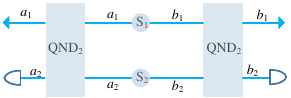}
\caption{The EPP based on ideal sources with the QND$_2$ \cite{EPP6}. The QND$_2$ is shown in Fig. 6 where $\theta  = \pi$.}
\end{center}
\end{figure}

For the four-photon state, i.e., $(a_{1H}^\dag b_{1H}^\dag  + a_{1V}^\dag b_{1V}^\dag  + a_{2H}^\dag b_{2H}^\dag  + a_{2V}^\dag b_{2V}^\dag )^2|0\rangle$, after photons passing through the QND$_1$, we have
\begin{eqnarray}\label{Fourphotonright}
&&(a_{1H}^\dag b_{1H}^\dag  + a_{2V}^\dag b_{2V}^\dag )^2|0\rangle |\alpha {e^{i2\theta }}{\rangle _a}|\alpha {e^{i2\theta }}{\rangle _b}+ (a_{1V}^\dag b_{1V}^\dag \nonumber\\
&&\!+\! a_{2H}^\dag b_{2H}^\dag )^2|0\rangle |\alpha {e^{i2{\theta ^\prime }}}{\rangle _a}|\alpha {e^{i2{\theta ^\prime }}}{\rangle _b}\!+\!2(a_{1H}^\dag b_{1H}^\dag\!+\!a_{2V}^\dag b_{2V}^\dag )\nonumber\\
&&(a_{1V}^\dag b_{1V}^\dag  + a_{2H}^\dag b_{2H}^\dag )|0\rangle |\alpha {e^{i{(\theta+\theta ^\prime) }}}{\rangle _a}|\alpha {e^{i{(\theta+\theta ^\prime) }}}{\rangle _b}.
\end{eqnarray}
It shows that Alice and Bob carry the same phase shift, i.e., $2\theta$, $2\theta^\prime$ or $\theta+\theta^\prime$, corresponding to ${\Omega _1}=(a_{1H}^\dag b_{1H}^\dag  + a_{2V}^\dag b_{2V}^\dag )^2|0\rangle$, ${\Omega _2}=(a_{1V}^\dag b_{1V}^\dag  + a_{2H}^\dag b_{2H}^\dag )^2|0\rangle$, and ${\Omega _3}=(a_{1H}^\dag b_{1H}^\dag  + a_{2V}^\dag b_{2V}^\dag )(a_{1V}^\dag b_{1V}^\dag  + a_{2H}^\dag b_{2H}^\dag )|0\rangle$, separately. Both the states ${\Omega _1}$ and ${\Omega _2}$ are in the same output modes after the coupler and they can not be identified in the spatial modes, thereby discarding them. However, ${\Omega _3}$ indicates that one pair emits from the upper mode and the other one emits from the lower mode which can be retained for quantum communication.

If the bit-flip error occurs on one of two pairs, for example, the state becomes $(a_{1H}^\dag b_{1H}^\dag  + a_{1V}^\dag b_{1V}^\dag  + a_{2H}^\dag b_{2H}^\dag  + a_{2V}^\dag b_{2V}^\dag )(a_{1V}^\dag b_{1H}^\dag  + a_{1H}^\dag b_{1V}^\dag  + a_{2V}^\dag b_{2H}^\dag  + a_{2H}^\dag b_{2V}^\dag )|0\rangle$. After the QND$_1$, it yields
\begin{eqnarray}\label{Fourphotononebitflip}
&&(a_{1H}^\dag b_{1H}^\dag  \!+\! a_{2V}^\dag b_{2V}^\dag )[(a_{1V}^\dag b_{1H}^\dag  \! +\!  a_{2H}^\dag b_{2V}^\dag )|0\rangle |\alpha {e^{i(\theta  + {\theta ^\prime })}}{\rangle _a}\nonumber\\
&& \otimes |\alpha {e^{i2\theta }}{\rangle _b} \! +\!  (a_{1H}^\dag b_{1V}^\dag  \! +\!  a_{2V}^\dag b_{2H}^\dag )|0\rangle |\alpha {e^{i2\theta }}{\rangle _a} |\alpha {e^{i(\theta  + {\theta ^\prime })}}{\rangle _b}] \nonumber\\
&&+(a_{1V}^\dag b_{1V}^\dag  \! +\!  a_{2H}^\dag b_{2H}^\dag )[(a_{1V}^\dag b_{1H}^\dag  \! +\!  a_{2H}^\dag b_{2V}^\dag )|0\rangle |\alpha {e^{i2{\theta ^\prime }}}{\rangle _a}\nonumber\\
&&\otimes|\alpha {e^{i(\theta  + {\theta ^\prime })}}{\rangle _b} \! +\!  (a_{1H}^\dag b_{1V}^\dag  \! +\!  a_{2V}^\dag b_{2H}^\dag )|0\rangle|\alpha {e^{i(\theta  + {\theta ^\prime })}}{\rangle _a}\nonumber\\
&&\otimes|\alpha {e^{i2{\theta ^\prime }}}{\rangle _b}],
\end{eqnarray}
which clearly illustrates that the different phase shifts are obtained for Alice and Bob by the $X$ homodyne measurement. Hence, these items can be automatically eliminated. In addition, if the bit-flip error happens on two pairs, the state will evolve to $(a_{1V}^\dag b_{1H}^\dag  + a_{1H}^\dag b_{1V}^\dag  + a_{2V}^\dag b_{2H}^\dag  + a_{2H}^\dag b_{2V}^\dag )(a_{1V}^\dag b_{1H}^\dag  + a_{1H}^\dag b_{1V}^\dag  + a_{2V}^\dag b_{2H}^\dag  + a_{2H}^\dag b_{2V}^\dag )|0\rangle$. With the same principle, the QND$_1$ changes the state to
\begin{eqnarray}\label{Fourphotontwobitflip}
&&{(a_{1H}^\dag b_{1V}^\dag  \!+\! a_{2V}^\dag b_{2H}^\dag )^2}|0\rangle |\alpha {e^{i2\theta }}{\rangle _a}|\alpha {e^{i2{\theta ^\prime }}}{\rangle _b} \!+\! 2(a_{1H}^\dag b_{1V}^\dag\nonumber\\
&&+ a_{2V}^\dag b_{2H}^\dag )(a_{1V}^\dag b_{1H}^\dag  \!+\! a_{2H}^\dag b_{2V}^\dag )|0\rangle |\alpha {e^{i(\theta  \!+\! {\theta ^\prime })}}{\rangle _a}|\alpha {e^{i(\theta  \!+\! {\theta ^\prime })}}{\rangle _b}\nonumber\\
&&+ {(a_{1V}^\dag b_{1H}^\dag  \!+\! a_{2H}^\dag b_{2V}^\dag )^2}|0\rangle |\alpha {e^{i2{\theta ^\prime }}}{\rangle _a}|\alpha {e^{i2\theta }}{\rangle _b}.
\end{eqnarray}
Alice and Bob will discard the items when they obtain the different phase shifts. While, if they get the same phase shift such as $\theta+\theta^\prime$, the state $(a_{1H}^\dag b_{1V}^\dag+a_{2V}^\dag b_{2H}^\dag )(a_{1V}^\dag b_{1H}^\dag  \!+\! a_{2H}^\dag b_{2V}^\dag )|0\rangle $ will be retained. Hence, the fidelity of the resultant state is $\frac{{2{p} + {p^2}{F^2}}}{{2{p} + {p^2}[{F^2} + {{(1 - F)}^2}]}}$.

After the purification, the bit-flip error has been suppressed to some extent. Moreover, the preserved two-photon state can be distinguished from the four-photon state according to the phase shifts. Therefore, one can further improve the fidelity of the mixed state. Let's assume the mixed state has the same form as Eq. (\ref{PBSmixed}). The principle of this EPP is shown in Fig. 13 and the QND$_2$ shown in Fig. 7. If $\theta  = \pi$, $|\phi^+\rangle_{a_1b_1}\otimes|\phi^+\rangle_{a_2b_2}$ becomes
\begin{eqnarray}\label{idealtwo}
&&|{\phi ^ + }{\rangle _{{a_1}{b_1}}} \otimes |{\phi ^ + }{\rangle _{{a_2}{b_2}}} \to \frac{1}{2}{[{(|HH\rangle _{{a_1}{b_1}}}|HH\rangle _{{a_2}{b_2}}}\nonumber\\
&&+|VV{\rangle _{{a_1}{b_1}}}|VV{\rangle _{{a_2}{b_2}}})|\alpha {e^{i\pi }}{\rangle _a}|\alpha {e^{i\pi }}{\rangle _b}\nonumber\\
&&+|HH{\rangle _{{a_1}{b_1}}}|VV{\rangle _{{a_2}{b_2}}}|\alpha {e^{i2\pi }}{\rangle _a}|\alpha {e^{i2\pi }}{\rangle _b}\nonumber\\
&&+|VV{\rangle _{{a_1}{b_1}}}|HH{\rangle _{{a_2}{b_2}}}|\alpha {\rangle _a}|\alpha {\rangle _b})].
\end{eqnarray}
The phase shifts carried by Alice and Bob are $\pi$, which means that $|{\phi ^ + }{\rangle _{{a_1}{b_1}}}|{\phi ^ + }{\rangle _{{a_2}{b_2}}}$ collapses to $|HH\rangle _{{a_1}{b_1}}|HH\rangle _{{a_2}{b_2}}+|VV\rangle _{{a_1}{b_1}}|VV\rangle _{{a_2}{b_2}}$. If the phase shifts are $2\pi$, the state becomes $|HH\rangle _{{a_1}{b_1}}|VV\rangle _{{a_2}{b_2}}+|VV\rangle _{{a_1}{b_1}}|HH\rangle _{{a_2}{b_2}}$ and an additional operator ${\sigma _x}$ is essential to be performed on both photons in modes $a_2$ and $b_2$. Subsequently, the same method as \cite{EPP3} can be adopted in the next step. The same analysis can be carried out for the remaining items. Therefore, the fidelity of this EPP with the QND$_2$ is $F_1$, but the efficiency of this EPP is twice than that of the EPP in Ref. \cite{EPP3}.

\begin{figure}
\begin{center}
\includegraphics[scale=1.8]{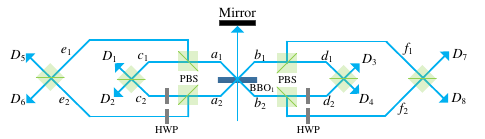}
\caption{The schematic diagram of one-step deterministic EPP using spatial mode entanglement \cite{EPP8}. After postselection, we can deterministically obtain the desired entanglement. To be specific, if $D_2$ and $D_4$ or $D_5$ and $D_7$ click, it means that the photons are in the polarization entanglement state $|\phi^+\rangle$. If the detectors $D_2$ and $D_7$ or $D_5$ and $D_4$ click, the photons are in $|\psi^+\rangle$ and an bit-flip operation is required on one of the photons to recover $|\phi^+\rangle$.}
\end{center}
\end{figure}

\subsection{One-step deterministic polarization EPP using spatial mode entanglement}
The concept of the deterministic entanglement purification was first proposed in Ref. \cite{EPP7}. This subsection will review the one-step deterministic polarization EPP using spatial mode entanglement \cite{EPP8}, which was experimentally realized \cite{EPPSB}. Moreover, robust time-bin entanglement can be also employed to achieve the deterministic polarization EPP \cite{Hyperentanglement11}, which was also experimentally realized \cite{EPPadd2}. As shown in Eq. (\ref{SPDC}), we assume the probability of generating one pair of photons from the SPDC source is small, i.e., $p \ll 1$. In this way, the higher-order items can be omitted. Additionally, let $\eta=1$ and $\omega=0$, the single-pair state can be given by
\begin{eqnarray}\label{SinglebySPDC}
|\Psi \rangle  = |\phi^ + {\rangle _{ab}}|\phi _s^ + {\rangle _{ab}}.
\end{eqnarray}
where $|\phi^+_s\rangle_{ab}$ is one of the four spatial mode Bell states
\begin{eqnarray}\label{spatialBellstates}
|\phi _s^ \pm {\rangle _{ab}} = \frac{1}{{\sqrt 2 }}(|{a_1}\rangle |{b_1}\rangle  \pm |{a_2}\rangle |{b_2}\rangle ),\nonumber\\
|\psi _s^ \pm {\rangle _{ab}} = \frac{1}{{\sqrt 2 }}(|{a_1}\rangle |{b_2}\rangle  \pm |{a_2}\rangle |{b_1}\rangle ).
\end{eqnarray}
During entanglement distribution over noisy channels, the quality of entanglement will unavoidably suffer from degradation. As discussed in Ref. \cite{EPP4}, the spatial entanglement keeps unchanged in noisy channels. Hence, the mixed state can be written as
\begin{eqnarray}\label{mixedspdc}
&&\rho_1  = {(F|{\phi ^ + }\rangle _{ab}}\langle {{\phi ^ + }}| + F_1|{\psi ^ + }{\rangle _{ab}}\langle {{\psi ^ + }}|\nonumber\\
&&+F_2|{\phi ^ - }{\rangle _{ab}}\langle {{\phi ^ - }}| + F_3|{\psi ^ - }{\rangle _{ab}}\langle {{\psi ^ - }}|)|\phi _s^ + {\rangle _{ab}}\langle {\phi _s^ + }|,
\end{eqnarray}

Let's first consider the item $|\phi^+\rangle_{ab}|\phi_s^ + \rangle _{ab}$. After passing through the PBSs and HWPs which makes transformation $|H\rangle  \leftrightarrow |V\rangle$, the state collapses to
\begin{eqnarray}\label{perfect}
|{\phi ^ + }{\rangle _{ab}}|\phi _s^ + {\rangle _{ab}} &\to& \frac{1}{2}({| H\rangle _{{c_1}}}{| H\rangle _{{d_1}}}+{| V\rangle _{{c_2}}}{| V\rangle _{{d_2}}} \nonumber\\
& +& {| V\rangle _{{e_1}}}{| V\rangle _{{f_1}}} + {| H\rangle _{{e_2}}}{| H\rangle _{{f_2}}}).
\end{eqnarray}
The components ${| H\rangle _{{c_1}}}{| H\rangle _{{d_1}}}$ and ${| V\rangle _{{c_2}}}{| V\rangle _{{d_2}}}$ make the coincidence detections on $D_2$ and $D_4$. While ${| V\rangle _{{e_1}}}{| V\rangle _{{f_1}}}$ and ${| H\rangle _{{e_2}}}{| H\rangle _{{f_2}}}$ make the coincidence detections on $D_5$ and $D_7$. It means that the desired state $|\phi^+\rangle$ is obtained.

If the bit-flip error occurs on $|\phi^+\rangle_{ab}$ with the probability of $F_1$ and after the PBSs and HWPs, the state $|\psi^+\rangle_{ab}|\phi^+_s\rangle_{ab}$ becomes
\begin{eqnarray}\label{bitflip}
|{\psi ^ + }\rangle_{ab} |\phi _s^ + \rangle_{ab}  &\to& \frac{1}{2}{(|H\rangle _{{c_1}}}|V{\rangle _{{f_1}}} + |V{\rangle _{{c_2}}}|H{\rangle _{{f_2}}}\nonumber\\
&+&|V{\rangle _{{e_1}}}|H{\rangle _{{d_1}}} + |H{\rangle _{{e_2}}}|V{\rangle _{{d_2}}}).
\end{eqnarray}
The components ${|H\rangle _{{c_1}}}|V{\rangle _{{f_1}}}$ and $|V{\rangle _{{c_2}}}|H{\rangle _{{f_2}}}$ make the coincidence detections on $D_2$ and $D_7$. The other remaining components ${|V\rangle _{{e_1}}}|H{\rangle _{{d_1}}}$ and $|H{\rangle _{{e_2}}}|V{\rangle _{{d_2}}}$ make the coincidence detections on $D_4$ and $D_5$. This projects the $|{\psi ^ + }\rangle_{ab} |\phi _s^ + \rangle_{ab}$ into $|\psi^+\rangle_{ab}$. Hence, we need to operate an additional bit-flip operation $\sigma_x$ on one of the photons to convert $|\psi^+\rangle_{ab}$ to $|\phi^+\rangle_{ab}$.

Similarly, if the phase-flip error occurs and after the PBSs and HWPs, the state will evolve to
\begin{eqnarray}\label{phaseflip}
|{\phi ^ -}\rangle_{ab} |\phi _s^ + \rangle_{ab}  &\to& \frac{1}{2}{(|H\rangle _{{c_1}}}|H{\rangle _{{d_1}}} + |V{\rangle _{{c_2}}}|V{\rangle _{{d_2}}}\nonumber\\
&-&|V{\rangle _{{e_1}}}|V{\rangle _{{f_1}}} - |H{\rangle _{{e_2}}}|H{\rangle _{{f_2}}}).
\end{eqnarray}
Obviously, the same result as Eq. (\ref{perfect}) can be obtained after postselection. Additionally, the state $|\psi^-\rangle_{ab}|\phi^+_s\rangle_{ab}$ has the same result as Eq. (\ref{bitflip}). Thus, the phase-flip error can be directly corrected. Moreover, the mixed state entangled in polarization is not essential for this deterministic EPP and it just requires to satisfy $F+F_1+F_2+F_3=1$. To elaborate, if the mixed state is
\begin{eqnarray}\label{noentanglement}
&&\rho_2  \!=\! {(F|HH\rangle _{ab}}{\langle HH| \!+\! {F_1}|VV\rangle _{ab}}\langle VV|\nonumber\\
&&\!+\!{F_2}|HV{\rangle _{ab}}{\langle HV| \!+\! {F_3}|VH\rangle _{ab}}\langle VH|)|\phi _s^+ {\rangle _{ab}}\langle \phi _s^ + |.
\end{eqnarray}
The two photons of the item $|HH{\rangle _{ab}}|\phi _s^ + {\rangle _{ab}}$ emit from the output modes $D_2$ and $D_4$, indicating that the state evolves to $|\phi^ + \rangle $. Additionally, if the measurement results are $D_5$ and $D_7$, it makes the state $|VV{\rangle _{ab}}|\phi _s^ + {\rangle _{ab}}$ project into $|\phi^ + \rangle $. Similarly, the components $|HV{\rangle _{ab}}|\phi _s^ + {\rangle _{ab}}$ ($|VH{\rangle _{ab}}|\phi _s^ + {\rangle _{ab}}$) will make the coincidence detectors $D_2$ and $D_7$ ($D_4$ and $D_5$) separately register one photon, which requires a bit-flip operation on one photon to transform $|\psi^ + \rangle$ to $|\phi ^ + \rangle$. In this way, the deterministic EPP has been carried out without requiring entanglement in polarization. This is different from the conventional EPPs \cite{EPP1,EPP2,EPP3,EPP4}, in which the initial fidelity should be larger than 0.5.

\subsection{Single-copy high-efficient EPP using hyperentanglement}
\begin{figure}
\begin{center}
\includegraphics[scale=2.0]{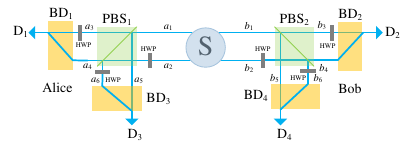}
\caption{The schematic diagram of the high efficient EPP using hyperentanglement \cite{EPP26}. The hyperentanglement source $S$ generates one pair of state $|\phi^+\rangle_{ab}|\phi^+_s\rangle_{ab}$. During the entanglement distribution, the hyperentangled state will be polluted because of the unavoidable noise, which will make the polarization entanglement and spatial entanglement evolve to the mixed state, separately. In addition, the beam displacer (BD) couples $| H\rangle$ and $| V\rangle$ into the same output mode. Only the two output modes $D_1D_2$ or $D_3D_4$ respectively contain one photon means a successful purification.}
\end{center}
\end{figure}
This subsection will focus on the single-copy high-efficient EPP using hyperentanglement after distributing the hyperentangled state over noisy channels \cite{EPP26}. In this EPP, if output modes $D_1$ and $D_2$ or $D_3$ and $D_4$ respectively contain one photon, it means a successful purification. To be specific, the source as shown in Fig. 15 generates one pair of hyperentanglement written as
\begin{eqnarray}\label{onehyperentanglement}
|\Phi\rangle=|\phi^+\rangle_{ab}\otimes|\phi^+_s\rangle_{ab},
\end{eqnarray}
After distributing photons to Alice and Bob, the state of Eq. (\ref{onehyperentanglement}) becomes
\begin{eqnarray}\label{Mixedstate}
\rho_3  &=& [{F_p}| {{\phi ^ + }}\rangle _{ab}\langle {{\phi ^ + }}| + (1 - {F_p})| {{\psi ^ + }}\rangle _{ab}\langle {{\psi ^ + }}|]\nonumber\\
 &\otimes& [{F_s}| {{\phi ^ +_s }}\rangle _{ab}\langle {{\phi ^ +_s }}| + (1 - {F_s})| {{\psi ^ +_s }}\rangle _{ab}\langle {{\psi ^ +_s }}|],
\end{eqnarray}
which can be described as the mixture of four pure states. The system is in the state $|{{\phi ^ + }}\rangle _{ab}|{{\phi ^ +_s }}\rangle _{ab}$, $|{{\psi ^ + }}\rangle _{ab}|{{\phi ^ +_s }}\rangle _{ab}$, $|{{\phi ^ + }}\rangle _{ab}|{{\psi ^ +_s }}\rangle _{ab}$, and $|{{\psi ^ + }}\rangle _{ab}|{{\psi ^ +_s }}\rangle _{ab}$ with the probability of $F_pF_s$, $(1-F_p)F_s$, $F_p(1-F_s)$, and $(1-F_p)(1-F_s)$, respectively. We first consider the item $|{{\phi ^ + }}\rangle _{ab}|{{\phi ^ +_s }}\rangle _{ab}$. After the actions of the PBSs and HWPs on photons, we have
\begin{eqnarray}\label{Firstitemevolve}
|{\phi ^ + }\rangle _{ab}|{\phi ^ +_s }\rangle _{ab} &\to& \frac{1}{2}{(|V\rangle _{{a_3}}}|V{\rangle _{{b_3}}} + |H{\rangle _{{a_4}}}|H{\rangle _{{b_4}}}\nonumber\\
 &+& |V{\rangle _{{a_5}}}|V{\rangle _{{b_5}}} + |H{\rangle _{{a_6}}}|H{\rangle _{{b_6}}}).
\end{eqnarray}
Using the beam displacers (BDs) to couple the photons with the different polarizations from the different input modes into the same output mode, thus the state of Eq. (\ref{Firstitemevolve}) collapses to
\begin{eqnarray}\label{FirstitemevolveD1D2}
|{\phi ^ + }\rangle _{{D_i}{D_{i+1}}} = \frac{1}{{\sqrt 2 }}{(|H\rangle _{{D_i}}}|H{\rangle _{{D_{i+1}}}} + |V{\rangle _{{D_i}}}|V{\rangle _{{D_{i+1}}}}),
\end{eqnarray}
where $i=1, 3$.
Similarly, the item $|{\psi ^ + }\rangle _{ab}|{\psi ^ +_s }\rangle _{ab}$ makes two photons emit from modes $D_1$ and $D_2$ or $D_3$ and $D_4$, which yields
\begin{eqnarray}\label{FouritemevolveD1D2}
|{\psi ^ + }\rangle _{{D_i}{D_{i+1}}} = \frac{1}{{\sqrt 2 }}{(|H\rangle _{{D_i}}}|V{\rangle _{{D_{i+1}}}} + |V{\rangle _{{D_i}}}|H{\rangle _{{D_{i+1}}}}).
\end{eqnarray}
Moreover, the other remaining states such as $|{\psi ^ + }\rangle _{ab}|{\phi ^ +_s }\rangle _{ab}$ and $|{\phi ^ + }\rangle _{ab}|{\psi ^ + _s }\rangle _{ab}$ result in each of output modes $D_1$ and $D_4$ or $D_2$ and $D_3$ containing one photon. For example, after the PBSs and HWPs, the state $|{\psi ^ + }\rangle _{ab}|{\phi ^ + _s}\rangle _{ab}$ becomes
\begin{eqnarray}\label{Twoitemevolve}
|{\psi ^ +}\rangle _{ab}|{\phi ^ +_s }\rangle _{ab} &\to& \frac{1}{2}{(|V\rangle _{{a_3}}}|V{\rangle _{{b_5}}} + |H{\rangle _{{a_6}}}|H{\rangle _{{b_4}}}\nonumber\\
 &+& |V{\rangle _{{a_5}}}|V{\rangle _{{b_3}}} + |H{\rangle _{{a_4}}}|H{\rangle _{{b_6}}}).
\end{eqnarray}
After passing through the BDs, it further evolves to
\begin{eqnarray}\label{FouritemevolveD1D4}
|{\psi ^ +}\rangle _{{D_1}{D_4}} = \frac{1}{{\sqrt 2 }}{(|H\rangle _{{D_1}}}|V{\rangle _{{D_4}}} + |V{\rangle _{{D_1}}}|H{\rangle _{{D_4}}}),
\end{eqnarray}
or
\begin{eqnarray}\label{FouritemevolveD2D3}
|{\psi ^ +}\rangle _{{D_2}{D_3}} = \frac{1}{{\sqrt 2 }}{(|H\rangle _{{D_2}}}|V{\rangle _{{D_3}}} + |V{\rangle _{{D_2}}}|H{\rangle _{{D_3}}}).
\end{eqnarray}
Obviously, it fails to satisfy our selection criterion thereby washing out it. The similar analysis can be done for $|{\phi ^ + }\rangle _{ab}|{\psi ^ +_s }\rangle _{ab}$. Consequently, the new mixed state can be obtained as
\begin{eqnarray}\label{newmixedD1D2}
{\rho _n} = {F_n}| {{\phi ^ +}}\rangle _{ab}\langle {{\phi ^ +}}| + (1 - {F_n})| {{\psi ^ +}}\rangle _{ab}\langle {{\psi ^ +}}|,
\end{eqnarray}
where
\begin{eqnarray}\label{newfidelity}
{F_n} = \frac{{{F_p}{F_s}}}{{{F_p}{F_s} + (1 - {F_p})(1 - {F_s})}}.
\end{eqnarray}
If both $F_p$ and $F_s$ are larger than 0.5, we can obtain ${F_n} > \max ({F_p},{F_s})$. According to the mentioned above, this EPP just uses single pair hyperentanglement to distill the high-fidelity mixed state entangled in polarization. Moreover, the purified entanglement can be further employed to improve the key rate for the entanglement-based QKD \cite{EPP26}.
\begin{figure}[!h]
\begin{center}
\includegraphics[scale=2.1]{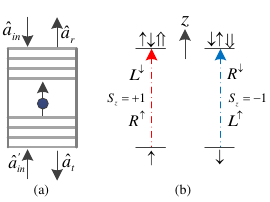}
\caption{(a) A double-sided QD cavity system, where the QD is in the center of a double-sided cavity \cite{Hyperentanglement12}. (b) The spin-based optical transitions of a negatively charged exciton $X^-$ with circularly polarized photons \cite{Hyperentanglement12}.}
\end{center}
\end{figure}

\subsection{The HEPP for nonlocal hyperentanglement systems}
This subsection introduces the HEPP for a nonlocal system entangled in polarization and spatial mode \cite{Hyperentanglement12}, which is different from the polarization EPPs consuming entanglement in other DOFs \cite{EPP4,EPP6,EPP26,Hyperentanglement8}. The authors employed the property of giant optical circular birefringence of a double-sided QD cavity system to construct parity-check gates and QSJMs to realize two-step HEPP. As shown in Fig. 16(a), a negatively charged exciton $X^-$ will be created provided that one puts a single electron into a QD. According to the rule of the Pauli's exclusion, it also shows spin-dependent transitions with circularly polarized lights, which is illustrated in Fig. 16(b).

\begin{figure}
\begin{center}
\includegraphics[scale=1.8]{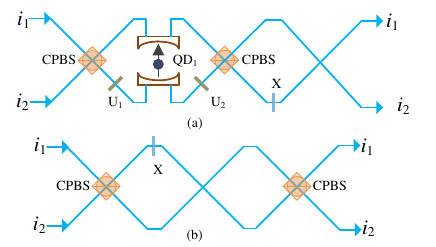}
\caption{(a) One QD, two CPBSs and two phase-flip operations ${{\rm{U}}_{1(2)}} =  -| R\rangle \langle R | - | L \rangle \langle L|$ as well as a bit-flip operation X for polarization make up of a QSJM \cite{Hyperentanglement12}. (b) The setup is used to swap the polarization state and spatial mode of a photon \cite{Hyperentanglement12}. The CPBS totally transmits the photon in polarization $|R\rangle$ and reflects $|L\rangle$.}
\end{center}
\end{figure}

The Heisenberg equations of motion for the cavity field operator ${\hat a}$ and $X^-$ dipole operator ${{\hat \sigma }_ - }$ can be employed to denote the input-output optical property of the double sided QD-cavity system, yielding
\begin{eqnarray}\label{newfidelity}
&&\frac{{d\hat a}}{{dt}} \!=\! -\! [i({w_c} \!-\! w) \!+\! \kappa  \!+\! \frac{{{\kappa _s}}}{2}]\hat a \!-\! g{{\hat \sigma }_ - }\!-\! \sqrt \kappa  {{\hat a}_{in}} \!-\! \sqrt \kappa  \hat a_{in}^\prime,\nonumber\\
&&\frac{{d{{\hat \sigma }_ - }}}{{dt}} \!=\! - [i({w_{{X^ - }}} \!-\! w) \!+\! \frac{\gamma }{2}]{{\hat \sigma }_ - } \!-\! g{{\hat \sigma }_z}\hat a,
\end{eqnarray}
where ${w_c}$, $w$, and ${w_{{X^ - }}}$ separately represent the frequencies of cavity, input photon, and the transition of $X^-$. The coefficient $g$ means the coupling strength of the exciton $X^-$ with negative charge and the cavity. $\frac{\gamma }{2}$ and $\frac{{{\kappa _s}}}{2}$ respectively denote the decay rates of exciton $X^-$ and the cavity field mode to the leakage mode. In addition, the two input operators are represented by ${{\hat a}_{in}}$ and $\hat a_{in}^\prime$. If the weak excitation occurs, one can obtain the expression of the transmission as
\begin{eqnarray}\label{transmission}
t(w) \!=\! \frac{{ - \kappa [i({w_{{X^ - }}} \!-\! w) \!+\! \frac{\gamma }{2}]}}{{[i({w_{{X^ - }}} \!-\! w) \!+\! \frac{\gamma }{2}][i({w_c} \!-\! w) \!+\! \kappa  \!+\! \frac{{{\kappa _s}}}{2}] \!+\! {g^2}}}.
\end{eqnarray}
Additionally, the reflection can be given by $r(w)=1+t(w)$. While if the strong-coupling and the resonant case satisfy at the same time, one can obtain $| r| \to 1$, $| r_0| \to 0$, $| t| \to 0$ and $| t_0| \to 1$. Due to the fact that the photonic circular polarization is determined by the direction of propagation, the photon is in state $|{R^ \uparrow }\rangle$ ($|{R^ \downarrow }\rangle$) or $|{L^ \downarrow }\rangle$ ($|{L^ \uparrow }\rangle$) simultaneously, which is shown in Fig. 16(b). Accordingly, the transmission and reflection of the photon polarization can be given by
\begin{eqnarray}\label{rule}
&&|{R^ \uparrow },{i_2}, \uparrow \rangle  \!\to\! |{L^ \downarrow },{i_2}, \uparrow \rangle ,{\rm{       }}|{L^ \downarrow },{i_1}, \uparrow \rangle  \!\to\! |{R^ \uparrow },{i_1}, \uparrow \rangle ,\nonumber\\
&&|{R^ \uparrow },{i_2}, \downarrow \rangle  \!\to\!  - |{R^ \uparrow },{i_2}, \downarrow \rangle ,{\rm{  }}|{L^ \downarrow },{i_1}, \downarrow \rangle  \!\to\!  - |{L^ \downarrow },{i_2}, \downarrow \rangle ,\nonumber\\
&&|{R^ \downarrow },{i_1}, \uparrow \rangle  \!\to\!  - |{R^ \downarrow },{i_2}, \uparrow \rangle ,{\rm{  }}|{L^ \uparrow },{i_2}, \uparrow \rangle  \!\to\!  - |{L^ \uparrow },{i_1}, \uparrow \rangle ,\nonumber\\
&&|{R^ \downarrow },{i_1}, \downarrow \rangle  \!\to\! |{L^ \uparrow },{i_1}, \downarrow \rangle ,{\rm{       }}|{L^ \uparrow },{i_2}, \downarrow \rangle  \!\to\!  - |{R^ \downarrow },{i_2}, \downarrow \rangle .
\end{eqnarray}
where $i=a, b$.

\begin{figure}
\begin{center}
\includegraphics[scale=1.8]{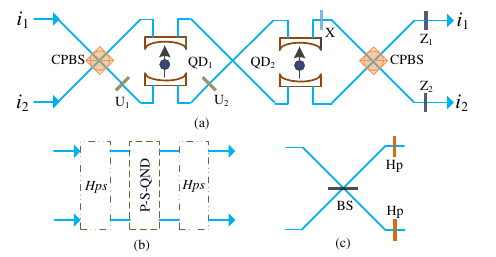}
\caption{(a) Two QDs, three CPBSs, two phase-flip operations ${{\rm{U}}_{1(2)}} =  -| R\rangle \langle R | - | L \rangle \langle L|$ and a bit-flip operation X for polarization as well as two operations ${{\rm{Z}}_{1(2)}} =  -| R\rangle \langle R | + | L \rangle \langle L|$ make up of a P-S-QND. \cite{Hyperentanglement12}. (b) The polarization-spatial parity-check QND is composed of two $H_{ps}$s and one P-S-QND \cite{Hyperentanglement12}. (c) The schematic diagram of the $H_{ps}$ \cite{Hyperentanglement12}. The BS is a 50:50 beam-splitter. H$_p$ is a Hadamard operation which effects $| R\rangle  \to \frac{1}{{\sqrt 2 }}(| R\rangle  + | L \rangle )$ and $| L\rangle  \to \frac{1}{{\sqrt 2 }}(| R\rangle  - | L \rangle )$.}
\end{center}
\end{figure}

The QSJM is utilized to transfer the polarization state of the photon $A$ to that of the photon $B$ while it remains the spatial mode of $B$ unchanged. The principle of this QSJM is shown in Fig. 17(a) which is composed of the double-side QD-cavity. Suppose that the initial states of $A$ and $B$ are
\begin{eqnarray}\label{initialstate}
|{\phi _A}\rangle  &=& ({\alpha _1}|R\rangle  + {\alpha _2}|L\rangle {)_A}({\gamma _1}|{a_1}\rangle  + {\gamma _2}|{a_2}\rangle ),\nonumber\\
|{\phi _B}\rangle  &=& ({\beta _1}|R\rangle  + {\beta _2}|L\rangle {)_B}({\delta _1}|{b_1}\rangle  + {\delta _2}|{b_2}\rangle ).
\end{eqnarray}
Additionally, we consider that the electron spin is in $| + {\rangle _e} = \frac{1}{{\sqrt 2 }}(| \uparrow \rangle  + | \downarrow \rangle {)_e}$. Thus, after passing through the setup as in Fig. 17(a), the state $| + {\rangle _e}|{\phi _A}\rangle$ becomes a new state which can be described as follows. If the polarization state of the photon $A$ is in $|R\rangle$, the electron spin is in ${\alpha _1}| \uparrow \rangle  + {\alpha _2}| \downarrow \rangle $. Contrarily, the electron spin is in ${\alpha _2}| \uparrow \rangle  + {\alpha _1}| \downarrow \rangle $ provided that the polarization state of $A$ is in $|L\rangle$. For the two cases, the spatial mode remains unchanged. To illustrate the principle of the QSJM, we only consider the case that the photon $A$ is in state $|R\rangle$, which yields ${\alpha _1}| \uparrow \rangle  + {\alpha _2}| \downarrow \rangle$. Subsequently, let the photon $B$ enter the setup as Fig. 17(a) after the action of the Hadamard operation on an electron spin, one can obtain
\begin{eqnarray}\label{initialstate1}
|{\phi _{Be}}{\rangle _1} &=& \frac{1}{{\sqrt 2 }}{[\alpha _1^\prime | \uparrow \rangle _e}({\beta _1}|R\rangle  + {\beta _2}|L\rangle {)_B} + \alpha _2^\prime | \downarrow {\rangle _e} \nonumber \\
  &&({\beta _2}|R\rangle  + {\beta _1}|L\rangle {)_B}]({\delta _1}|{b_1}\rangle  + {\delta _2}|{b_2}\rangle ),
\end{eqnarray}
where $\alpha _1^\prime  = {\alpha _1} + {\alpha _2}$ and $\alpha _2^\prime  = {\alpha _1} - {\alpha _2}$.
Subsequently, we perform the Hadamard operations on polarization of the photon $B$ and the electron spin passes through the circuit shown in Fig. 17(a). Finally, after adding another Hadamard operation on the electron spin, one can obtain
\begin{eqnarray}\label{finalstate1}
|{\phi _{Be}}{\rangle _2} &=& \frac{1}{{\sqrt 2 }}{[{\beta _1}| \uparrow \rangle _e}({\alpha _1}|R\rangle  + {\alpha _2}|L\rangle {)_B} +{\beta _2}| \downarrow {\rangle _e} \nonumber\\
&&({\alpha _1}|R\rangle  - {\alpha _2}|L\rangle {)_B}]({\delta _1}|{b_1}\rangle  + {\delta _2}|{b_2}\rangle ).
\end{eqnarray}
The polarization state of the photon $A$ has been transferred to that of the photon $B$ without changing the spatial mode of $B$ by measuring the electron spin state. Moreover, if we combine the quantum circuit Fig. 17(b) with Fig. 17(a) in sequence, we can achieve the purpose that the spatial mode state of the photon $A$ can be transferred to polarization state of the photon $B$ without distributing the spatial mode of $B$.

\begin{figure}
\begin{center}
\includegraphics[scale=2.2]{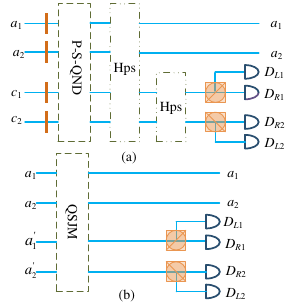}
\caption{The schematic diagram of the (a) first step HEPP \cite{Hyperentanglement12} and (b) second HEPP \cite{Hyperentanglement12}.}
\end{center}
\end{figure}
The polarization-spatial phase check QND (P-S-QND) is constructed in Fig. 18(a). This quantum circuit can unambiguously identify the relative phase for the hyperentangled state. There are 16 hyperentangled states in polarization and spatial mode, which can be written as $|{\phi _{kl}}{\rangle _{AB}} = |{\phi _k}\rangle _{AB}^P|{\phi _l}\rangle _{AB}^S$ with $k,l=1,2,3,4$. Here, $|{\phi _k}\rangle _{AB}^P$ and $|{\phi _l}\rangle _{AB}^S$ are described as
\begin{eqnarray}\label{Bell}
|{\phi _1}\rangle _{AB}^P &=& \frac{1}{{\sqrt 2 }}(|RR\rangle  + |LL\rangle {)_{AB}},\nonumber\\
|{\phi _2}\rangle _{AB}^P &=& \frac{1}{{\sqrt 2 }}(|RR\rangle  - |LL\rangle {)_{AB}},\nonumber\\
|{\phi _3}\rangle _{AB}^P &=& \frac{1}{{\sqrt 2 }}(|RL\rangle  + |LR\rangle {)_{AB}},\nonumber\\
|{\phi _4}\rangle _{AB}^P &=& \frac{1}{{\sqrt 2 }}(|RL\rangle  - |LR\rangle {)_{AB}},\nonumber\\
|{\phi _1}\rangle _{AB}^s &=& \frac{1}{{\sqrt 2 }}(|{a_1}{b_1}\rangle  + |{a_2}{b_2}\rangle ),\nonumber\\
|{\phi _2}\rangle _{AB}^s &=& \frac{1}{{\sqrt 2 }}(|{a_1}{b_1}\rangle  - |{a_2}{b_2}\rangle ),\nonumber\\
|{\phi _3}\rangle _{AB}^s &=& \frac{1}{{\sqrt 2 }}(|{a_1}{b_2}\rangle  + |{a_2}{b_1}\rangle ),\nonumber\\
|{\phi _4}\rangle _{AB}^s &=& \frac{1}{{\sqrt 2 }}(|{a_1}{b_2}\rangle  - |{a_2}{b_1}\rangle ).
\end{eqnarray}
We consider that the electron spins of the QD$_1$ and QD$_2$ are in $|+\rangle_{e_1}$ and $|+\rangle_{e_2}$. We let the photons $A$ and $B$ enter circuit as shown in Fig. 18(a) in turn followed by measuring the electron spin in the basis $|  \pm \rangle $. If the measurement outcome in $e_1$ is $|+\rangle$, the polarization state is in $|{\phi _1}\rangle _{AB}^P$ or $|{\phi _3}\rangle _{AB}^P$. If the measurement outcome of $e_1$ is in $|-\rangle$,  the polarization state is in $|{\phi _2}\rangle _{AB}^P$ or $|{\phi _4}\rangle _{AB}^P$. Similarly, the spatial mode entanglement is in $|{\phi _1}\rangle _{AB}^S$ or $|{\phi _3}\rangle _{AB}^S$ if $e_2$ is in $|-\rangle$. While if $e_2$ is in $|+\rangle$, the spatial mode entanglement is in $|{\phi _2}\rangle _{AB}^S$ or $|{\phi _4}\rangle _{AB}^S$. Thus, it can be concluded as
\begin{eqnarray}\label{hypercnot}
|{\phi _{{k_1}{l_1}}}{\rangle _{AB}}| + {\rangle _{{e_1}}}| + {\rangle _{{e_2}}} \to |{\phi _{{k_1}{l_1}}}{\rangle _{AB}}| + {\rangle _{{e_1}}}| - {\rangle _{{e_2}}},\nonumber\\
|{\phi _{{k_1}{l_2}}}{\rangle _{AB}}| + {\rangle _{{e_1}}}| + {\rangle _{{e_2}}} \to |{\phi _{{k_1}{l_2}}}{\rangle _{AB}}| + {\rangle _{{e_1}}}| + {\rangle _{{e_2}}},\nonumber\\
|{\phi _{{k_2}{l_1}}}{\rangle _{AB}}| + {\rangle _{{e_1}}}| + {\rangle _{{e_2}}} \to |{\phi _{{k_2}{l_1}}}{\rangle _{AB}}| - {\rangle _{{e_1}}}| - {\rangle _{{e_2}}},\nonumber\\
|{\phi _{{k_2}{l_2}}}{\rangle _{AB}}| + {\rangle _{{e_1}}}| + {\rangle _{{e_2}}} \to |{\phi _{{k_2}{l_2}}}{\rangle _{AB}}| - {\rangle _{{e_1}}}| + {\rangle _{{e_2}}}.
\end{eqnarray}
To further distinguish the parity of polarization and spatial mode, the Hadamard operations are required to act on both the polarization and spatial modes. So far, the QSJM and P-S-QND have been briefly illustrated. In the following, we will explain the HEPP.  Suppose that the initial mixed hyperentangled state is
\begin{eqnarray}\label{hypermixed}
\rho_4  &=& {[{F_p}|{\phi _1}\rangle ^P}\langle {{\phi _1}}| + (1 - {F_p})|{\phi _3}{\rangle ^P}\langle {{\phi _3}}|]\nonumber\\
&&\otimes{[{F_s}|{\phi _1}\rangle ^S}\langle {{\phi _1}}| + (1 - {F_s})|{\phi _2}{\rangle ^S}\langle {{\phi _2}}|].
\end{eqnarray}
The two noisy copies $\rho_{AB}$ and $\rho_{CD}$ have the same form as Eq. (\ref{hypermixed}). Consequently, the whole system consists of 16 hyperentangled states. After the actions of Fig. 19(a) and measuring the electron spin, we can unambiguously acquire the information on parity of photon pairs in the polarization and spatial mode. There are four different cases which are discussed in the following parts.

The first case is that the polarization and spatial mode have the same parity mode for the photon pairs $AC$ and $BD$. In this way, the state $|\phi_1\rangle_{AB}^P\otimes|\phi_1\rangle_{AB}^P$ ($|\phi_1\rangle_{AB}^S\otimes|\phi_1\rangle_{AB}^S$) and $|\phi_3\rangle_{AB}^P\otimes|\phi_3\rangle_{AB}^P$ ($|\phi_3\rangle_{AB}^S\otimes|\phi_3\rangle_{AB}^S$) can be discriminated from $|\phi_1\rangle_{AB}^P\otimes|\phi_3\rangle_{AB}^P$ ($|\phi_1\rangle_{AB}^S\otimes|\phi_3\rangle_{AB}^S$) and $|\phi_3\rangle_{AB}^P\otimes|\phi_1\rangle_{AB}^P$ ($|\phi_3\rangle_{AB}^S\otimes|\phi_1\rangle_{AB}^S$). By measuring the electron spin state, they evolves to
\begin{eqnarray}\label{sameparity}
&&|{\Phi _1}{\rangle _P} = \frac{1}{{\sqrt 2 }}(|RRRR\rangle  + |LLLL\rangle {)_{ABCD}},\nonumber\\
&&|{\Phi _2}{\rangle _P} = \frac{1}{{\sqrt 2 }}(|RRLL\rangle  + |LLRR\rangle {)_{ABCD}},\nonumber\\
&&|{\Phi _3}{\rangle _P} = \frac{1}{{\sqrt 2 }}(|RLRL\rangle  + |LRLR\rangle {)_{ABCD}},\nonumber\\
&&|{\Phi _4}{\rangle _P} = \frac{1}{{\sqrt 2 }}(|RLLR\rangle  + |LRRL\rangle {)_{ABCD}},\nonumber\\
&&|{\Phi _1}{\rangle _S} = \frac{1}{{\sqrt 2 }}(|{a_1}{b_1}{c_1}{d_1}\rangle  + |{a_2}{b_2}{c_2}{d_2}\rangle ),\nonumber\\
&&|{\Phi _2}{\rangle _S} = \frac{1}{{\sqrt 2 }}(|{a_1}{b_1}{c_2}{d_2}\rangle  + |{a_2}{b_2}{c_1}{d_1}\rangle ),\nonumber\\
&&|{\Phi _3}{\rangle _S} = \frac{1}{{\sqrt 2 }}(|{a_1}{b_2}{c_1}{d_2}\rangle  + |{a_2}{b_1}{c_2}{d_1}\rangle ),\nonumber\\
&&|{\Phi _4}{\rangle _S} = \frac{1}{{\sqrt 2 }}(|{a_1}{b_2}{c_2}{d_1}\rangle  + |{a_2}{b_1}{c_1}{d_2}\rangle ).
\end{eqnarray}
Obviously, $|{\Phi _2}{\rangle _P}$, $|{\Phi _4}{\rangle _P}$, $|{\Phi _2}{\rangle _S}$, and $|{\Phi _4}{\rangle _S}$ can be transformed to $|{\Phi _1}{\rangle _P}$, $|{\Phi _3}{\rangle _P}$, $|{\Phi _1}{\rangle _S}$, and $|{\Phi _3}{\rangle _S}$ when one adds the bit-flip operations on the photons $C$ and $D$. We perform the Hadamard operations on both polarization and spatial mode of $C$ and $D$ for states $|{\Phi _1}{\rangle _P}$, $|{\Phi _3}{\rangle _P}$, $|{\Phi _1}{\rangle _S}$, and $|{\Phi _3}{\rangle _S}$. According to the measurement results of the detectors, we pick out the even cases for polarization and spatial mode. Thus, we can obtain a new mixed state entangled in polarization and spatial mode with $F_p^\prime = \frac{{F_p^2}}{{F_p^2 + {{(1 - {F_p})}^2}}}$ and $F_s^\prime  = \frac{{F_s^2}}{{F_s^2 + {{(1 - {F_s})}^2}}}$.

The second case is that both the polarization and spatial mode have the different parity modes for the photon pairs $AC$ and $BD$. For example, $|\phi_1\rangle_{AB}^P\otimes|\phi_3\rangle_{AB}^P$ ($|\phi_1\rangle_{AB}^S\otimes|\phi_2\rangle_{AB}^S$) and $|\phi_3\rangle_{AB}^P\otimes|\phi_1\rangle_{AB}^P$ ($|\phi_2\rangle_{AB}^S\otimes|\phi_1\rangle_{AB}^S$) can be discriminated from $|\phi_1\rangle_{AB}^P\otimes|\phi_1\rangle_{AB}^P$ ($|\phi_1\rangle_{AB}^S\otimes|\phi_1\rangle_{AB}^S$) and $|\phi_3\rangle_{AB}^P\otimes|\phi_3\rangle_{AB}^P$ ($|\phi_2\rangle_{AB}^S\otimes|\phi_2\rangle_{AB}^S$). By measuring the electron spin using the P-S-QND, we can obtain
\begin{eqnarray}\label{differentparity}
&&|{\Phi _5}{\rangle _P} = \frac{1}{{\sqrt 2 }}(|RRRL\rangle + |LLLR\rangle {)_{ABCD}},\nonumber\\
&&|{\Phi _6}{\rangle _P} = \frac{1}{{\sqrt 2 }}(|RRLR\rangle + |LLRL\rangle {)_{ABCD}},\nonumber\\
&&|{\Phi _7}{\rangle _P} = \frac{1}{{\sqrt 2 }}(|RLRR\rangle + |LRLL\rangle {)_{ABCD}},\nonumber\\
&&|{\Phi _8}{\rangle _P} = \frac{1}{{\sqrt 2 }}(|RLLL\rangle + |LRRR\rangle {)_{ABCD}},\nonumber\\
&&|{\Phi _5}{\rangle _S} = \frac{1}{{\sqrt 2 }}(|{a_1}{b_1}{c_1}{d_2}\rangle  + |{a_2}{b_2}{c_2}{d_1}\rangle ),\nonumber\\
&&|{\Phi _6}{\rangle _S} = \frac{1}{{\sqrt 2 }}(|{a_1}{b_1}{c_2}{d_1}\rangle  + |{a_2}{b_2}{c_1}{d_2}\rangle ),\nonumber\\
&&|{\Phi _7}{\rangle _S} = \frac{1}{{\sqrt 2 }}(|{a_1}{b_2}{c_1}{d_1}\rangle  + |{a_2}{b_1}{c_2}{d_2}\rangle ),\nonumber\\
&&|{\Phi _8}{\rangle _S} = \frac{1}{{\sqrt 2 }}(|{a_1}{b_2}{c_2}{d_2}\rangle  + |{a_2}{b_1}{c_1}{d_1}\rangle ).
\end{eqnarray}
With the same principle, this case can be discarded resulted from the fact that two parties fail to distinguish the bit-flip error in polarization and phase-flip error in spatial mode.

The third case is that both the photon pairs $AC$ and $BD$ are in the same polarization parity mode and different spatial parity modes. In other words, the polarization states $|\phi_1\rangle_{AB}^P\otimes|\phi_1\rangle_{AB}^P$ ($|\phi_1\rangle_{AB}^S\otimes|\phi_2\rangle_{AB}^S$) and $|\phi_3\rangle_{AB}^P\otimes|\phi_3\rangle_{AB}^P$ ($|\phi_2\rangle_{AB}^S\otimes|\phi_1\rangle_{AB}^S$) can be discriminated from $|\phi_1\rangle_{AB}^P\otimes|\phi_3\rangle_{AB}^P$ ($|\phi_1\rangle_{AB}^S\otimes|\phi_1\rangle_{AB}^S$) and $|\phi_3\rangle_{AB}^P\otimes|\phi_1\rangle_{AB}^P$ ($|\phi_2\rangle_{AB}^S\otimes|\phi_2\rangle_{AB}^S$). By measuring the electron spin state using the P-S-QNDs, the polarization state collapses to $|{\Phi _1}{\rangle _P}$ ($|{\Phi _2}{\rangle _P}$) and $|{\Phi _3}{\rangle _P}$ ($|{\Phi _4}{\rangle _P}$) and the spatial mode becomes $|{\Phi _5}{\rangle _S}$, $|{\Phi _6}{\rangle _S}$, $|{\Phi _7}{\rangle _S}$ and $|{\Phi _8}{\rangle _S}$. Obviously, the polarization state can be reused to distill high-fidelity entanglement. In this way, one can employ the QSJM to purify in a second step.

\begin{figure}
\begin{center}
\includegraphics[scale=0.5]{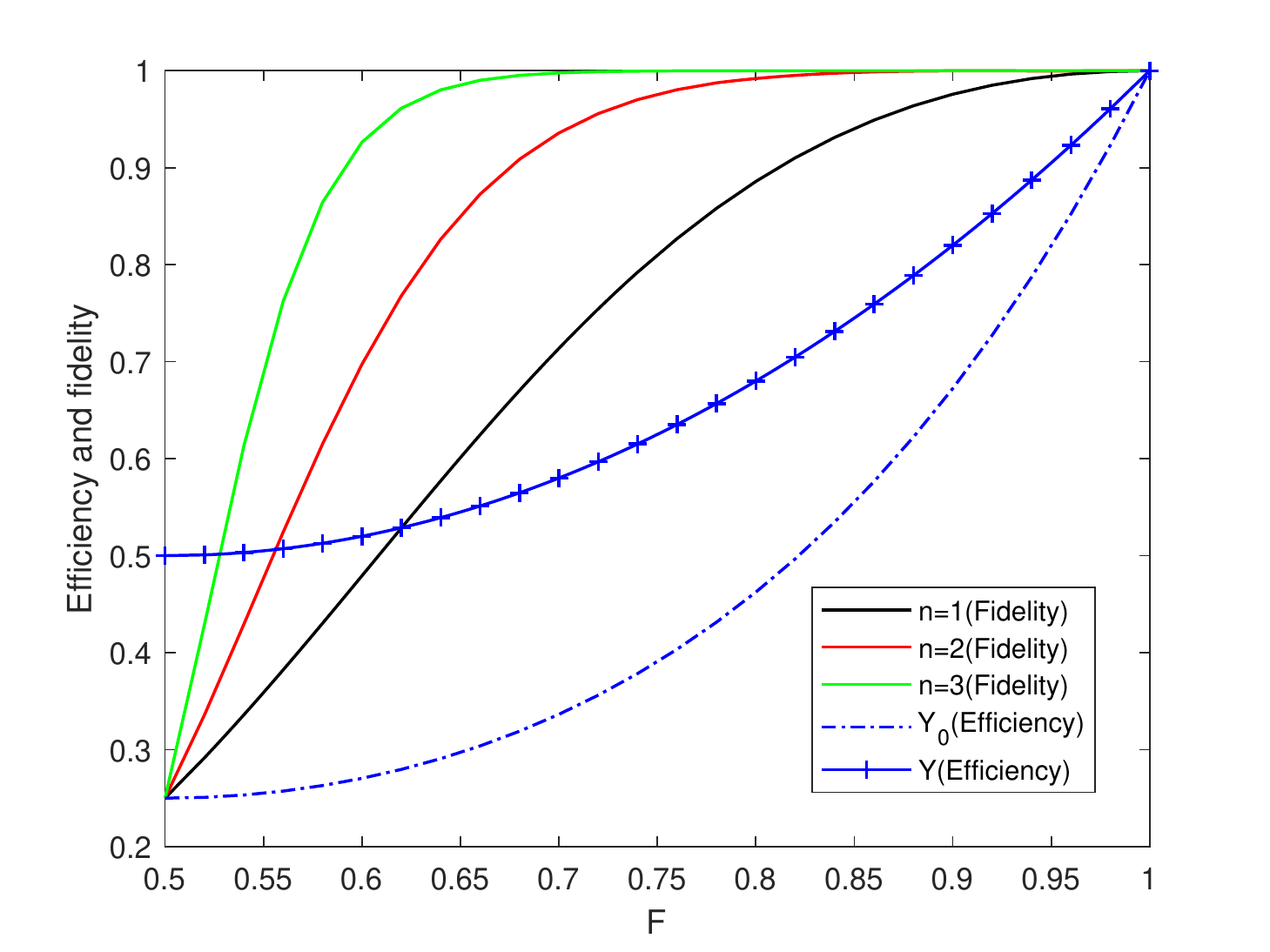}
\caption{The fidelity and efficiency of the HEPP \cite{Hyperentanglement12} with the QSJM and the conventional HEPP without the QSJM versus the initial fidelity.}
\end{center}
\end{figure}

Finally, the last case is that both the photon pairs $AC$ and $BD$ are in the different polarization parity modes and same spatial parity mode. For instance, $|\phi_1\rangle_{AB}^P\otimes|\phi_3\rangle_{AB}^P$ ($|\phi_1\rangle_{AB}^S\otimes|\phi_1\rangle_{AB}^S$) and $|\phi_3\rangle_{AB}^P\otimes|\phi_1\rangle_{AB}^P$ ($|\phi_2\rangle_{AB}^S\otimes|\phi_2\rangle_{AB}^S$) can be discriminated from $|\phi_1\rangle_{AB}^P\otimes|\phi_1\rangle_{AB}^P$ ($|\phi_1\rangle_{AB}^S\otimes|\phi_2\rangle_{AB}^S$) and $|\phi_3\rangle_{AB}^P\otimes|\phi_3\rangle_{AB}^P$ ($|\phi_2\rangle_{AB}^S\otimes|\phi_1\rangle_{AB}^S$). The measurement outcomes of electron spin states using the P-S-QNDs make the polarization states evolve to $|{\Phi _5}{\rangle _P}$ ($|{\Phi _6}{\rangle _P}$) and $|{\Phi _7}{\rangle _P}$ ($|{\Phi _8}{\rangle _P}$) and the spatial-mode states become $|{\Phi _1}{\rangle _S}$ ($|{\Phi _2}{\rangle _S}$) and $|{\Phi _3}{\rangle _S}$ ($|{\Phi _4}{\rangle _S}$). In this circumstance, the QSJM is required in a second step.

For the third case and last case, four totally identical photon pairs $AB$, $CD$, ${A^\prime }{B^\prime }$, and ${C^\prime }{D^\prime }$ are essential for the second step purification. Here, the photons $A$, ${A^\prime }$, $C$, and ${C^\prime }$ are hold by Alice. The remaining photons belong to Bob. Similar to the first step, the same actions are required to perform for $ABCD$ and ${A^\prime }{B^\prime }{C^\prime }{D^\prime }$. If the measurement outcomes of electron spin states make the four photon pairs project to the third and last (last and third) cases, the polarization (spatial-mode) state of $AB$ and the spatial-mode (polarization) state of ${A^\prime }{B^\prime }$ are combined into an output of photon pair with the QSJM. In this way, the similar method can be applied to the other cases.

Consequently, adding a Hadamard operation on the spatial-mode, we can obtain a new mixed state $\rho_{AB}^{\prime}$ according to the two-step HEPP. Here,
\begin{eqnarray}\label{new}
\rho _{AB}^\prime  &=& [F_p^\prime| {{\phi _1}}\rangle _{AB}^P\langle {{\phi _1}}| + (1 - F_p^\prime )| {{\phi _3}}\rangle _{AB}^P\langle {{\phi _3}} |]\nonumber\\
 &\otimes& [F_s^\prime | {{\phi _1}}\rangle _{AB}^S\langle {{\phi _1}}| + (1 - F_s^\prime )| {{\phi _3}}\rangle _{AB}^S\langle {{\phi _3}}|],
\end{eqnarray}
with the fidelity of $F_p^{\prime}F_s^{\prime}$ which is shown in Fig. 20.  Moreover, the efficiency of this HEPP with the QSJM is $F_s^2 + {(1 - {F_s})^2}$ which is larger than that ($[F_p^2 + {(1 - {F_p})^2}][F_s^2 + {(1 - {F_s})^2}]$) in the convention HEPP without the QSJM.

\section{The MBEPP}
In 1998, the concept of the quantum repeater was proposed to realize the long-distance quantum communication \cite{repeater1}. However, the building blocks of quantum repeaters are needed to work with sufficiently high accuracy \cite{MBQR}, which makes quantum repeaters hard to implement in current technology. To this end, the measurement-based quantum repeater (MBQR) was proposed \cite{MBQR}, where only simple BSAs \cite{BSM1} were required to couple the resource states to input noisy states instead of performing coherent two-qubit gates. In the MBQR, the core ingredient is the MBEPP \cite{HighErrorTolerance2}. The research findings show that the MBEPP can tolerate much more local noise than the conventional EPPs based on the CNOT gates or similar logical operations \cite{HighErrorTolerance2}. Subsequently, the feasibility of the MBEPP in linear optics with the SPDC sources was investigated \cite{EPP25}. Recently, the authors further studied the MBEPP for entangled coherent states \cite{EPPaddyan} and investigated the MBEPP for logical qubit entanglement with imperfect QND under the photon loss \cite{EPPPRAyan}.

\subsection{The principle of the universal MBEPP}
\begin{figure}
\begin{center}
\includegraphics[scale=2.5]{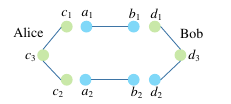}
\caption{The schematic diagram of the MBEPP \cite{HighErrorTolerance2,MBQR}. The standard BSAs \cite{BSM1} are required to operate on the photons in modes $c_1$ and $a_1$, $c_2$ and $a_2$, $b_1$ and $d_1$, and $b_2$ and $d_2$. The resource states entangled in modes $c_1c_2c_3$ and $d_1d_2d_3$ are prepared off-line in a probabilistic way. As pointed out in Refs. \cite{HighErrorTolerance2,MBQR}, this MBEPP can purify $m$ noisy copies with $m+1$-particle resource states each round of purification.}
\end{center}
\end{figure}

\begin{table}[!htbp]
\centering
\caption{The measurement outcomes and corresponding operations performed on output states. The first column denotes the measurement outcomes on ${g_1}{a_1}$, ${g_2}{a_2}$, ${h_1}{b_1}$, and ${h_2}{b_2}$. The second column represents the parity of ``$-$", i.e., $|\phi^{-}_n\rangle_{{c_1}{a_1}}|\phi^{-}_n\rangle_{{c_2}{a_2}}|\phi^{-}_n\rangle_{{d_1}{b_1}}|\phi^{-}_n\rangle_{{d_2}{b_2}}$ means even ``$-$" case and $|\psi^{+}_n\rangle_{{c_1}{a_1}}|\phi^{-}_n\rangle_{{c_2}{a_2}}|\psi^{-}_n\rangle_{{d_1}{b_1}}|\phi^{-}_n\rangle_{{d_2}{b_2}}$ means odd ``$-$" case. The third column means that the additional operations (${\sigma _{x1}} = |0\rangle\langle 1| - |1\rangle\langle 0|$) are required to be operated.}
\begin{tabular}{|c|c|c|c|c|}
\hline
\multicolumn{2}{|l|}{$|\phi_n\rangle_{{c_1}{a_1}}|\phi_n\rangle_{{c_2}{a_2}}|\phi_n\rangle_{{d_1}{b_1}}|\phi_n\rangle_{{d_2}{b_2}}$} & \multicolumn{2}{l|}{even ``$-$"} & I \\
\multicolumn{2}{|l|}{$|\phi_n\rangle_{{c_1}{a_1}}|\psi_n\rangle_{{c_2}{a_2}}|\phi_n\rangle_{{d_1}{b_1}}|\psi_n\rangle_{{d_2}{b_2}}$} & \multicolumn{2}{l|}{}                  &                   \\ \cline{3-5}
\multicolumn{2}{|l|}{$|\psi_n\rangle_{{c_1}{a_1}}|\phi_n\rangle_{{c_2}{a_2}}|\psi_n\rangle_{{d_1}{b_1}}|\phi_n\rangle_{{d_2}{b_2}}$} & \multicolumn{2}{l|}{odd ``$-$"} &  $\sigma_{z1}$\\
\multicolumn{2}{|l|}{$|\psi_n\rangle_{{c_1}{a_1}}|\psi_n\rangle_{{c_2}{a_2}}|\psi_n\rangle_{{d_1}{b_1}}|\psi_n\rangle_{{d_2}{b_2}}$} & \multicolumn{2}{l|}{}                  &                   \\
\hline
\multicolumn{2}{|l|}{$|\phi_n\rangle_{{c_1}{a_1}}|\phi_n\rangle_{{c_2}{a_2}}|\psi_n\rangle_{{d_1}{b_1}}|\psi_n\rangle_{{d_2}{b_2}}$} & \multicolumn{2}{l|}{even ``$-$"} &  $\sigma_{x1}$\\
\multicolumn{2}{|l|}{$|\phi_n\rangle_{{c_1}{a_1}}|\psi_n\rangle_{{c_2}{a_2}}|\psi_n\rangle_{{d_1}{b_1}}|\phi_n\rangle_{{d_2}{b_2}}$} & \multicolumn{2}{l|}{}                  &                   \\ \cline{3-5}
\multicolumn{2}{|l|}{$|\psi_n\rangle_{{c_1}{a_1}}|\phi_n\rangle_{{c_2}{a_2}}|\phi_n\rangle_{{d_1}{b_1}}|\psi_n\rangle_{{d_2}{b_2}}$} & \multicolumn{2}{l|}{odd ``$-$"} & $\sigma_{x1}\sigma_{z1}$ \\
\multicolumn{2}{|l|}{$|\psi_n\rangle_{{c_1}{a_1}}|\psi_n\rangle_{{c_2}{a_2}}|\phi_n\rangle_{{d_1}{b_1}}|\phi_n\rangle_{{d_2}{b_2}}$} & \multicolumn{2}{l|}{}                  &                   \\ \hline
\end{tabular}
\end{table}

As shown in Fig. 21, two pairs of three-particle resource states prepared off-line in a probabilistic way are coupled to the noisy pairs via the standard BSAs \cite{BSM1}. The three-particle resource state can be given by $|\phi _0^ + \rangle =\frac{1}{{\sqrt 2 }}(|000\rangle  + |111\rangle )$, which entangles in modes $c_1c_2c_3$ and $d_1d_2d_3$, respectively. It needs to be pointed out that the MBEPP can purify $n$ noisy copies with $n+1$-particle resource states each time. Additionally, the resource states can be 2D cluster states. For simplicity, we only consider the case that two noisy pairs and resource states are used to distill high-fidelity of the Bell states from low-quality of entanglement. The noisy pairs $\rho_{a_1b_1}$ (it is in the state $|\phi _n^ + {\rangle _{{a_1}{b_1}}}$ ($|\psi _n^ + {\rangle _{{a_1}{b_1}}}$) with the probability of $F$ ($1-F$)) and $\rho_{a_2b_2}$ (it is in the state $|\phi _n^ + {\rangle _{{a_2}{b_2}}}$ ($|\psi _n^ + {\rangle _{{a_2}{b_2}}}$) with the probability of $F$ ($1-F$)). Thus, the whole system $\rho_{a_1b_1}\otimes\rho_{a_2b_2}$ combines with the two resource states $|\phi _0^ + \rangle_{c_1c_2c_3}$ and $|\phi _0^ + \rangle_{d_1d_2d_3}$ can be described as follows. It is in the state $|\phi _0^ + \rangle_{c_1c_2c_3}\otimes|\phi _0^ + \rangle_{d_1d_2d_3}\otimes|\phi_n^+\rangle_{a_1b_1}\otimes|\phi_n^+\rangle_{a_2b_2}$ with the probability of $F^2$. It is in the state $|\phi _0^ + \rangle_{c_1c_2c_3}\otimes|\phi _0^ + \rangle_{d_1d_2d_3}\otimes|\psi_n^+\rangle_{a_1b_1}\otimes|\psi_n^+\rangle_{a_2b_2}$ with the probability of $(1-F)^2$. It is in the state $|\phi _0^ + \rangle_{c_1c_2c_3}\otimes|\phi _0^ + \rangle_{d_1d_2d_3}\otimes|\psi_n^+(\phi_n^+)\rangle_{a_1b_1}\otimes|\phi_n^+(\psi_n^+)\rangle_{a_2b_2}$ with the probability of $F(1-F)$. We merely consider the item $|\phi _0^ + \rangle_{c_1c_2c_3}\otimes|\phi _0^ + \rangle_{d_1d_2d_3}\otimes|\phi_n^+\rangle_{a_1b_1}\otimes|\phi_n^+\rangle_{a_2b_2}$ to illustrate the principle of the MBEPP, which is written as
\begin{eqnarray}\label{MBEPPright}
&&|\phi _0^ + \rangle _{{c_1}{c_2}{c_3}} \otimes |\phi _0^ + \rangle _{{d_1}{d_2}{d_3}}\otimes |{\phi_n ^ + }{\rangle _{{a_1}{b_1}}} \otimes |{\phi_n ^ + }{\rangle _{{a_2}{b_2}}}\nonumber\\
&=&\frac{1}{4}(|000\rangle  + |111\rangle {)_{{c_1}{c_2}{c_3}}}(|000\rangle  + |111\rangle {)_{{d_1}{d_2}{d_3}}}\nonumber\\
&\otimes& (|00\rangle  + |11\rangle {)_{{a_1}{b_1}}}(|00\rangle  + |11\rangle {)_{{a_2}{b_2}}}.
\end{eqnarray}
Then, one can perform the BSAs on the particles in modes $c_1a_1$, $c_2a_2$, $b_1d_1$, and $b_2d_2$. As discussed in Ref. \cite{MBQR}, the MBEPP is successful provided that the measurement outcomes at Alice and Bob are the same. To be specific, if one obtains the measurement outcomes such as $|\phi_n\rangle|\phi_n\rangle|\phi_n\rangle|\phi_n\rangle$, $|\phi_n\rangle|\psi_n\rangle|\phi_n\rangle|\psi_n\rangle$, $|\psi_n\rangle|\phi_n\rangle|\psi_n\rangle|\phi_n\rangle$ and $|\psi_n\rangle|\psi_n\rangle|\psi_n\rangle|\psi_n\rangle$ and we have even ``$-$" case, i.e., $|\phi^{-}_n\rangle|\phi^{-}_n\rangle|\phi^{-}_n\rangle|\phi^{-}_n\rangle$,  the state as Eq. (\ref{MBEPPright}) evolves to
\begin{eqnarray}\label{MBEPPrightevolvephi+}
|{\phi_n ^ + }{\rangle _{{c_3}{d_3}}} = \frac{1}{{\sqrt 2 }}{(|00\rangle _{{c_3}{d_3}}} + |11{\rangle _{{c_3}{d_3}}}).
\end{eqnarray}
If the odd ``$-$" case, i.e., $|\phi^{+}_n\rangle|\phi^{-}_n\rangle|\phi^{-}_n\rangle|\phi^{-}_n\rangle$, is obtained, it collapses to
\begin{eqnarray}\label{MBEPPrightevolvephi+}
|{\phi_n ^ - }{\rangle _{{c_3}{d_3}}} = \frac{1}{{\sqrt 2 }}{(|00\rangle _{{c_3}{d_3}}} - |11{\rangle _{{c_3}{d_3}}}),
\end{eqnarray}
and an unitary phase-flip operation ${\sigma _{z1}} = |0\rangle\langle 0| - |1\rangle\langle 1|$ should be performed on one of the particles. With the same principle, the cross combinations $|\phi_n^+\rangle_{a_1b_1}\otimes|\psi_n^+\rangle_{a_2b_2}$ and $|\psi_n^+\rangle_{a_1b_1}\otimes|\phi_n^+\rangle_{a_2b_2}$ can be washed out according to the measurement outcomes. Similarly, the state $|\phi _0^ + \rangle_{c_1c_2c_3}\otimes|\phi _0^ + \rangle_{d_1d_2d_3}\otimes|\psi_n^+\rangle_{a_1b_1}\otimes|\psi_n^+\rangle_{a_2b_2}$ collapses to $|{\psi _n^ + }{\rangle _{{c_3}{d_3}}}$ with or without additional operations on one of the photons according to the measurement outcomes. As a consequence, the final mixed state can be obtained as
\begin{eqnarray}\label{MBEPPnewmixed}
\rho_{{c_3}{d_3}} = F_1{| {{\phi_n ^ + }} \rangle _{c_3d_3}}\langle {{\phi_n ^ + }} | + (1-F_1){| {{\psi_n ^ + }}\rangle _{c_3d_3}}\langle {{\psi_n ^ + }}|.
\end{eqnarray}
The other cases and the corresponding unitary operations required to perform are detailed in Table I.

\subsection{The MBEPP in linear optics}
In the practical scenario, the entanglement is usually generated by the SPDC source. At first glance, the double-pair emissions emitted from the SPDC source may pose big challenges in both the preparation of resource states and entanglement purification.  As shown in Fig. 22, the pump pulse passes through a BBO crystal to generate entanglement as Eq. (\ref{entanglementbyspdc}). The bit-flip error may occur  with the probability of $1-F$ during the entanglement distribution, and the state as Eq. (\ref{entanglementbyspdc}) will evolve to the state as Eq. (\ref{mixedentanglementbyspdc}).

\begin{figure}
\begin{center}
\includegraphics[scale=0.6]{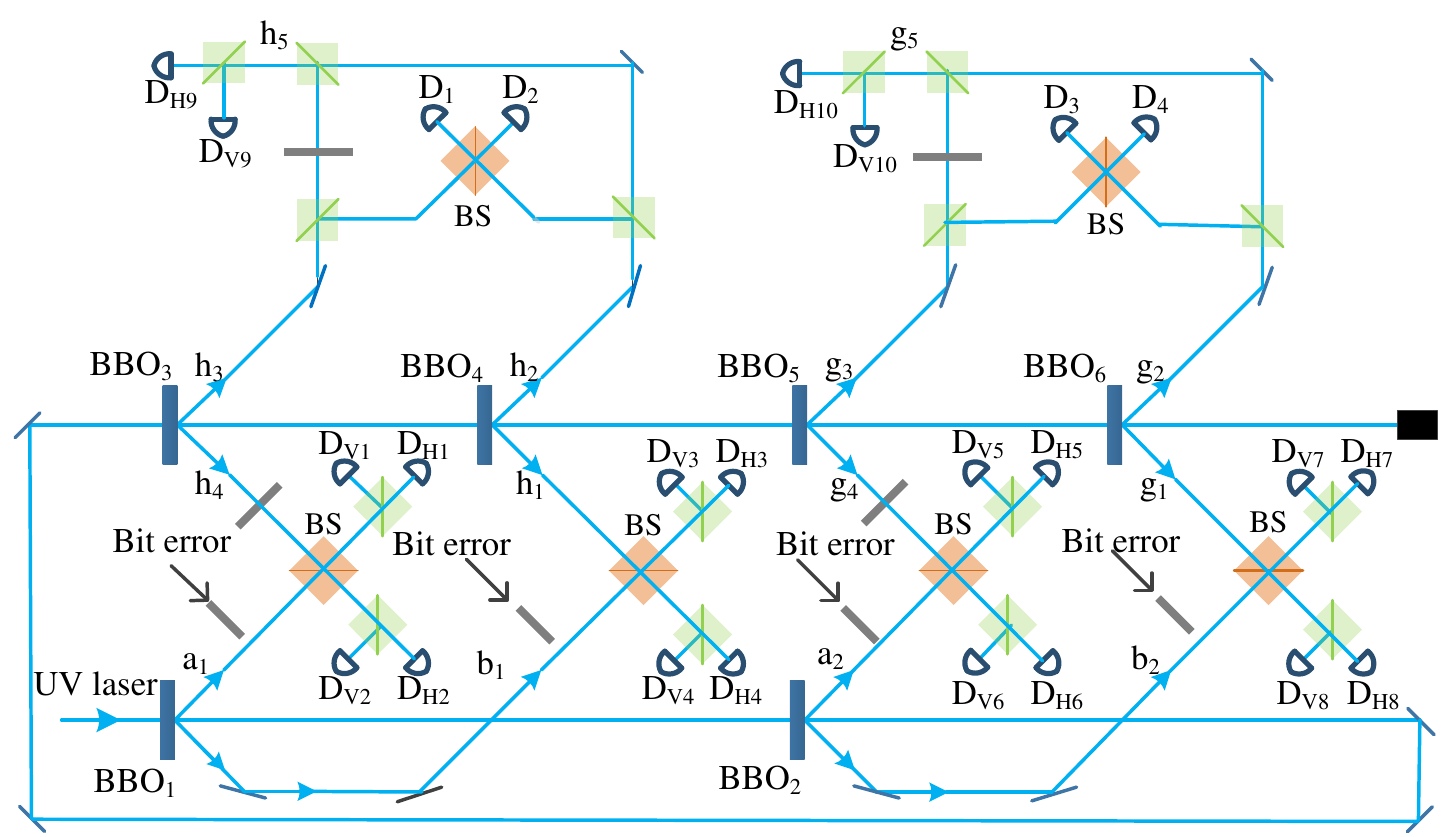}
\caption{The schematic diagram of the MBEPP in linear optics \cite{EPP25}. This MBEPP consists of three parts. The first is to use the BBO$_1$ and BBO$_2$ to produce two noisy copies. The second part is to produce three-photon resource state. Here, we use clicks on $D_1$ and $D_3$ to herald the success generation of resource state. The third part is the standard BSA \cite{BSM1}. We pick out ``twelve mode" case to herald a successful case for the MBEPP.}
\end{center}
\end{figure}

During the preparation of resource states, we employ the BBO$_3$ (BBO$_5$) and BBO$_4$ (BBO$_6$) to generate resource states \cite{EPP25}. Additionally, a click on detector $D_1$ ($D_3$) is used to herald a success generation of the resource state. Let the photons pass through the PBSs, BSs, and HWPs. The resource state generated by the BBO$_3$ and BBO$_4$ can be obtained as
\begin{eqnarray}\label{resourcestatebySPDC}
&&|{\rm{Res}}{\rangle _1} \!=\! \frac{{\sqrt p }}{{2\sqrt 2 }}{(|V\rangle _{{h_1}}} \!+\! |H{\rangle _{{h_4}}}) \!+\! \frac{p}{4}|V{\rangle _{{h_1}}}|H{\rangle _{{h_4}}}\nonumber\\
&&+\frac{p}{{\sqrt 2 }}{(|H\rangle _{{h_5}}}|H{\rangle _{{h_1}}}|V{\rangle _{{h_1}}} \!+\! |V{\rangle _{{h_5}}}|H{\rangle _{{h_4}}}|V{\rangle _{{h_4}}})\nonumber\\
&&+\frac{p}{4}{(|V\rangle _{{h_1}}}|V{\rangle _{{h_1}}} \!+\! |H{\rangle _{{h_4}}}|H{\rangle _{{h_4}}}) \!+\! \frac{p}{2}|{\rm{\phi_4^+}}{\rangle _{{h_1}{h_4}{h_5}}}.\nonumber\\
\end{eqnarray}
Similarly, the second resource state $|{\rm{Res}}{\rangle _2}$ has the same form as Eq. (\ref{resourcestatebySPDC}). Hence, the system $\rho_{a_1b_1}\otimes\rho_{a_2b_2}$ combined with $|{\rm{Res}}{\rangle _1}\otimes|{\rm{Res}}{\rangle _2}$ can be written as
\begin{eqnarray}\label{composite}
&&{\rho _{{a_1}{b_1}}} \otimes {\rho _{{a_2}{b_2}}} \otimes |{\rm{Res}}{\rangle _1} \otimes |{\rm{Res}}{\rangle _2}\nonumber\\
&=& {[F|{\Phi ^ + }\rangle _{{a_1}{b_1}}}{\langle {\Phi ^ + }| + (1 - F)|{\Psi ^ + }\rangle _{{a_1}{b_1}}}\langle {\Psi ^ + }|]\nonumber\\
&\otimes& {[F|{\Phi ^ + }\rangle _{{a_2}{b_2}}}{\langle {\Phi ^ + }| + (1 - F)|{\Psi ^ + }\rangle _{{a_2}{b_2}}}\langle {\Psi ^ + }|]\nonumber\\
&\otimes&{[\frac{{\sqrt p }}{{2\sqrt 2 }}{(|V\rangle _{{h_1}}} + |H\rangle _{{h_4}}}) + \frac{p}{4}|V{\rangle _{{h_1}}}|H{\rangle _{{h_4}}} + \frac{p}{{\sqrt 2 }}\nonumber\\
&\times& {(|H\rangle _{{h_5}}}|H{\rangle _{{h_1}}}|V{\rangle _{{h_1}}} + |V{\rangle _{{h_5}}}|H{\rangle _{{h_4}}}|V{\rangle _{{h_4}}}) + \frac{p}{4}\nonumber\\
&\times& {(|V\rangle _{{h_1}}}|V{\rangle _{{h_1}}} + |H{\rangle _{{h_4}}}|H{\rangle _{{h_4}}}) + \frac{p}{2}|{\rm{\phi_4^+}}{\rangle _{{h_1}{h_4}{h_5}}}]\nonumber\\
&\otimes& {[\frac{{\sqrt p }}{{2\sqrt 2 }}{(|V\rangle _{{g_1}}} + |H\rangle _{{g_4}}}) + \frac{p}{4}|V{\rangle _{{g_1}}}|H{\rangle _{{g_4}}} + \frac{p}{{\sqrt 2 }}\nonumber\\
&\times& {(|H\rangle _{{g_5}}}|H{\rangle _{{g_1}}}|V{\rangle _{{g_1}}} + |V{\rangle _{{g_5}}}|H{\rangle _{{g_4}}}|V{\rangle _{{g_4}}}) + \frac{p}{4}\nonumber\\
&\times& {(|V\rangle _{{g_1}}}|V{\rangle _{{g_1}}} + |H{\rangle _{{g_4}}}|H{\rangle _{{g_4}}}) + \frac{p}{2}|{\rm{\phi_4^+}}{\rangle _{{g_1}{g_4}{g_5}}}].
\end{eqnarray}

The successful operations of four BSAs and the preparations of resource states are crucial to realize the MBEPP. Thus, we pick out the ``twelve mode" case to herald a successful purification. To elaborate, the coincidence detections on $D_{H1}D_{V1}$, $D_{H3}D_{V3}$, $D_{H5}D_{V5}$, and $D_{H7}D_{V7}$ indicate that all measurement outcomes are the same such as $|\psi^+\rangle$. Moreover, the clicks on detectors $D_1$ and $D_3$ are used to herald the success preparation for resource states. Furthermore, the photons in modes $h_5$ and $g_5$ are required to be detected to indicate a successful MBEPP. Therefore, the ``twelve mode" case will yield a new mixed state with the higher fidelity compared to the initial one.

Here, we only consider the case that $\rho_{a_1b_1}\otimes\rho_{a_2b_2}$ is in the state $|\phi^+\rangle_{a_1b_1}\otimes|\phi^+\rangle_{a_2b_2}$ with the probability of $F^2p^2$ and one pair of resource states is in $|{\rm{\phi_4^+}}\rangle_{h_1h_4h_5}$ and the other is in $|V\rangle_{g_5}|H\rangle_{g_4}|V\rangle_{g_4}$. In this case, the state in Eq. (\ref{composite}) becomes
\begin{eqnarray}\label{composite1}
&&\frac{1}{{2\sqrt 2 }}(|HH\rangle  + |VV\rangle {)_{{a_1}{b_1}}} \otimes (|HH\rangle  + |VV\rangle {)_{{a_2}{b_2}}}\nonumber\\
&&\otimes(|HHH\rangle  + |VVV\rangle {)_{{h_1}{h_4}{h_5}}} \otimes |VHV{\rangle _{{g_5}{g_4}{g_4}}}.
\end{eqnarray}
Obviously, the mode $g_1$ is always absent of photon, which fails to satisfy the ``twelve mode'' case. The analysis for the other remaining components of Eq. (\ref{composite}) can be done with or without a little modification. As a result, only each of the noisy copies is a single pair and each of the resource states is $|\phi_4^+\rangle$ can make contributions to the MBEPP.

\subsection{The MBEPP for logical qubit entanglement}
This subsection reviews the MBEPP for logical qubit entanglement where the qubit is encoded in quantum parity code \cite{errorcorrection6,errorcorrection7}. It has three levels such as physical level, block level, and logical level. As a result, the logical Bell states are written as
\begin{eqnarray}\label{lbellstate}
{|{{\phi _{k,l}}}\rangle ^ * } = \frac{1}{{\sqrt 2 }}[{| {0,{\rm{ }}k}\rangle ^ * } + {( - 1)^l}{| {1,{\rm{ 1}} - k}\rangle ^ * }],
\end{eqnarray}
where ''$*$'', ''$k$'' and ''$l$'' ($k, l= 0, 1$) denote the corresponding level, the amplitude bit, and the phase bit. In this case, the higher level Bell states can be represented by the lower level Bell states, i.e.,
\begin{eqnarray}\label{lowerlevel}
&&{| {{\phi _{k,l}}}\rangle ^{(m)} } = \frac{1}{{\sqrt {{2^{m + 1}}} }}\sum\limits_{\overrightarrow r  \in {S_{l,m}}} {\mathop  \otimes \limits_{i = 1}^m | {{\phi _{k,{r_i}}}} \rangle },\nonumber\\
&&{| {{\phi _{k,l}}}\rangle ^{(n,m)} } = \frac{1}{{\sqrt {{2^{n + 1}}} }}\sum\limits_{\overrightarrow r  \in {S_{k,n}}} {\mathop  \otimes \limits_{i = 1}^n {{| {{\phi _{{r_i},l}}}\rangle }^{(m)}}},
\end{eqnarray}
where $m$ and $n$ respectively denote each block contains $m$ photons and each logical level is made up of $n$ blocks. Additionally, ${S_{l,m}} = \{ \overrightarrow r  \in {\{ 0,1\} ^m}|\sum\limits_{i = 1}^m {{r_i} \oplus 2 = l} \}$ and ${S_{k,n}} = \{ \vec r \in {\{ 0,1\} ^n}|\sum\limits_{i = 1}^n {{r_i} \oplus 2 = k} \}$.

\begin{figure}[t]
\includegraphics[scale=2]{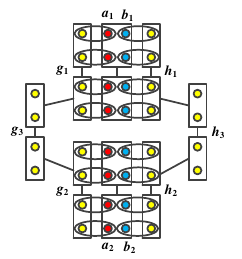}\caption{The schematic diagram of the MBEPP for logical Bell states with $n=m=2$ \cite{EPPPRAyan}. Each oval represents a BSA for the physical Bell states.  We use the yellow circles to represent the qubits of resource states. The blue and red circles respectively denote the qubits hold by Bob and Alice.  \label{logical_MBEPP}}
\end{figure}

\begin{figure}[t]
\includegraphics[scale=1]{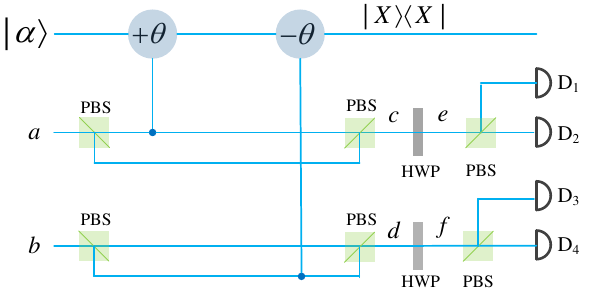}\caption{The schematic diagram of the BSA for the physical level with QND$_3$ \cite{EPPPRAyan}. Based on the phase shifts of $|\alpha\rangle$, the physical Bell states can be divided into two groups $({|{{\phi _{0,0}}}\rangle}, {|{{\phi _{0,1}}}\rangle})$ and $({|{{\phi _{1,0}}}\rangle}, {|{{\phi _{1,1}}}\rangle})$. Then, ${|{{\phi _{0,0}}}\rangle}$ (${|{{\phi _{1,0}}}\rangle}$) and ${|{{\phi _{0,1}}}\rangle}$ (${|{{\phi _{1,1}}}\rangle}$) can be identified according to the response of the photon detectors $D_{i}$ (i=1,2,3,4).
\label{QND}}
\end{figure}

We only take $n=m=2$ as an example to illustrate the principle of the MBEPP for logical qubit entanglement as shown in Fig. \ref{logical_MBEPP} \cite{EPPPRAyan}.  Each oval represents a BSA with the QND$_3$ in physical level depicted in Fig. \ref{QND}. The logical resource state is given by
\begin{eqnarray}\label{resource}
&&| {{\text{GHZ}}}\rangle_{{j_1}{j_2}{j_3}} \!\!=\!\! \frac{1}{{\sqrt 2 }}({| {{\text{000}}}\rangle ^{(n,m)}} \!+\! {| {{\text{111}}}\rangle ^{(n,m)}})_{{j_1}{j_2}{j_3}} ,
\end{eqnarray}
where $j \in \left\{ {g,h} \right\}$. Consequently, we can purify the noisy logical entanglement based on the measurement outcomes after performing the logical BSAs on the logical qubits $g_1a_1$, $g_2a_2$, $h_1b_1$, and $h_2b_2$. If all the photons remain intact during the distribution and the perfect QND$_3$ is available, the similar analysis as the MBEPP for the physical qubit can be adopted. Specifically, if the measurement outcomes $|{\phi _{{k_1},{l_1}}}\rangle _{{g_1}{a_1}}^{(n,m)}$, $|\phi_{k_2,l_2}\rangle_{g_2a_2}^{(n,m)}$, $|\phi_{k_3,l_3}\rangle_{h_1b_1}^{(n,m)}$, and $|\phi_{k_4,l_4}\rangle_{h_2b_2}^{(n,m)}$ satisfy the condition that
\begin{eqnarray}\label{r}
{k_{i + 2}} = {k_i} \oplus 2,\sum\limits_{m = 1}^4 {{l_m}}  \oplus 2 = 0.
\end{eqnarray}
The resultant state is
\begin{eqnarray}\label{finalmixed}
{\rho _{{g_3}{h_3}}} \!=\! {F_1}|{\phi _{0,0}}\rangle _{{g_3}{h_3}}^{(n,m)}\langle {\phi _{0,0}}| \!+\! (1 \!-\! {F_1})|{\phi _{1,0}}\rangle _{{g_3}{h_3}}^{(n,m)}\langle {\phi _{1,0}}|.
\end{eqnarray}
In addition, the new mixed state can be given by
\begin{eqnarray}\label{finalmixed2}
\rho _{{g_3}{h_3}}^Z \!=\! {F_1}{| {{\phi _{0,1}}}\rangle ^{(n,m)}}\langle {{\phi _{0,1}}}|\!+\!(1 \!\!-\!\! {F_1}){|{{\phi _{1,1}}} \rangle ^{(n,m)}}\langle {{\phi _{1,1}}}|,
\end{eqnarray}
under the condition that
\begin{eqnarray}\label{rule2}
{k_{i + 2}} = {k_i} \oplus 2,\sum\limits_{m = 1}^4 {{l_m}}  \oplus 2 = 1.
\end{eqnarray}
In this case, the additional phase-flip operations are needed to be done on the first photon and the fifth photon to transform $\rho _{{g_3}{h_3}}^Z$ to ${\rho _{{g_3}{h_3}}}$.
If the measurement outcomes are
\begin{eqnarray}\label{rule3}
{k_{i + 2}} = ({k_i}+1) \oplus 2,\sum\limits_{m = 1}^4 {{l_m}}  \oplus 2 = 0,
\end{eqnarray}
it yields a new mixed state as
\begin{eqnarray}\label{finalmixed3}
\rho _{{g_3}{h_3}}^X \!=\! {F_1}{| {{\phi _{1,0}}}\rangle ^{(n,m)}}\langle {{\phi _{1,0}}}|\!+\!(1 \!\!-\!\! {F_1}){|{{\phi _{0,0}}} \rangle ^{(n,m)}}\langle {{\phi _{0,0}}}|.
\end{eqnarray}
If the additional bit-flip operations are performed on the sixth photon and the eighth photon, we can obtain ${\rho _{{g_3}{h_3}}}$.
Finally, if the measurement outcomes satisfy the condition that
\begin{eqnarray}\label{rule4}
{k_{i + 2}} = ({k_i}+1) \oplus 2,\sum\limits_{m = 1}^4 {{l_m}}  \oplus 2 = 1,
\end{eqnarray}
the resultant state can be given by
\begin{eqnarray}\label{finalmixed4}
\rho _{{g_3}{h_3}}^{XZ} \!=\! {F_1}{| {{\phi _{1,1}}}\rangle ^{(n,m)}}\langle {{\phi _{1,1}}}|\!+\!(1 \!\!-\!\! {F_1}){|{{\phi _{0,1}}} \rangle ^{(n,m)}}\langle {{\phi _{0,1}}}|.
\end{eqnarray}
We can first transform the state $\rho _{{g_3}{h_3}}^{XZ}$ to the state $\rho _{{g_3}{h_3}}^X$ after the additional phase-flip operations on the first and fifth photons. After that, the bit-flip operations on the sixth and the eighth photons can be performed to obtain the state ${\rho _{{g_3}{h_3}}}$.

\begin{figure}[t]
\includegraphics[scale=0.5]{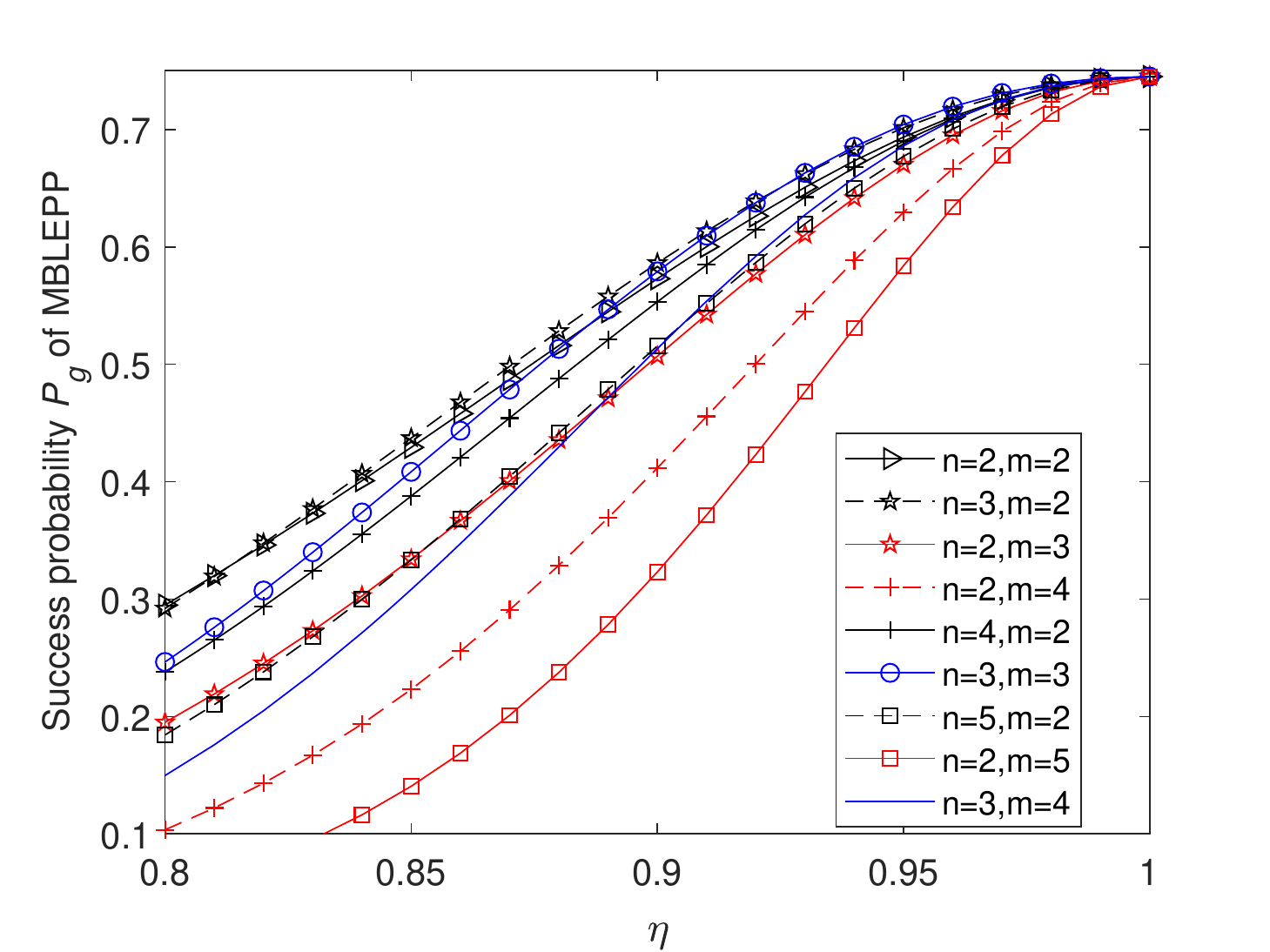}\caption{The success probability $P_{g}$ of the purification with the perfect QND$_3$ versus the photon transmission efficiency $\eta$ with $F=0.85$  \cite{EPPPRAyan}.  \label{success}}
\end{figure}

However, the photon loss is inevitable because of the noisy environment. As discussed in \cite{errorcorrection6,errorcorrection7}, the logical BSA can be successfully performed provided that at least one of the block is intact and each block contains at least one photon. In this way, one can successfully operate the MBEPP for logical qubit entanglement as long as the photon loss is less than the tolerance threshold of the QPC and the entanglement between different blocks still exists. Based on this conclusion, we can obtain the success probability of the MBEPP for logical qubit entanglement under the photon loss as
\begin{eqnarray}\label{totallsuccess}
{P_{g}} = \sum\limits_{j = 0}^{{n_t}} {\prod\limits_{s = 1}^4 {{P_{{j_s}}}} {E_j}} {(1 - \eta )^j}{\eta ^{4mn - j}}{P_F},
\end{eqnarray}
where $P_F = F^2 +(1 - F)^2$. If the purification is successful under the condition that two noisy copies totally lose $j$ photons, we have $E_j=1$. Contrarily, $E_j=0$. $P_{j_s}$ denotes the success probability of a logical BSA under the photon loss and $s$ ($s=1,2,3,4$) represents the $s$th logical BSA. Moreover, ${n_t}$ represents the loss tolerance threshold of $\rho_{a_1b_1}$ and $\rho_{a_2b_2}$. Fig. \ref{success} shows the success probability of the MBEPP for logical qubit entanglement versus the photon transmission efficiency $\eta$. It obviously illustrates that an increasing $\eta$ results in an enhancement on $P_g$.

\begin{figure}[t]
\includegraphics[scale=0.5]{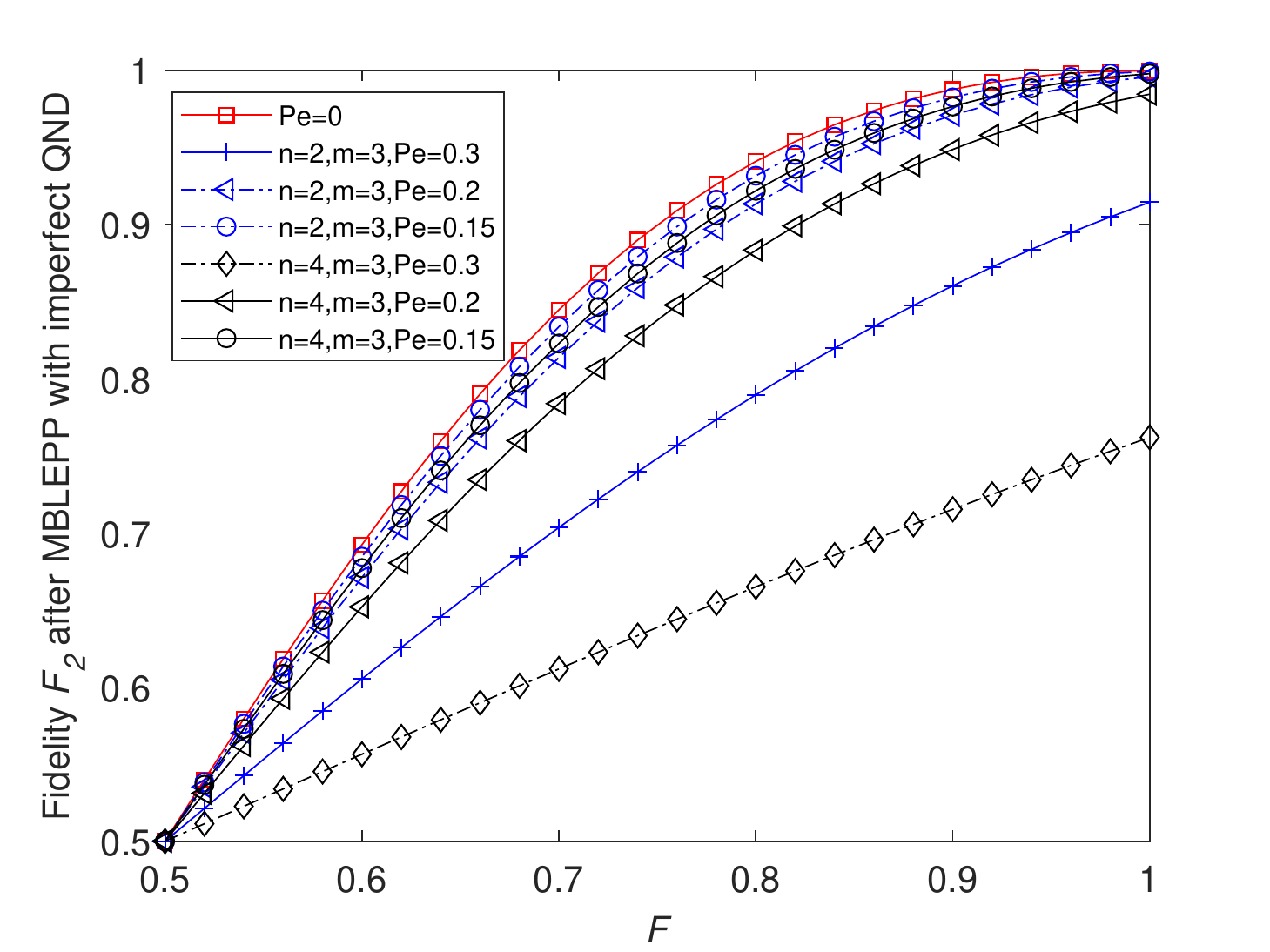}\caption{The fidelity $F_{2}$ after the purification with the imperfect QND$_3$ versus the initial fidelity $F$ for different $P_e$ \cite{EPPPRAyan}.  \label{Fidelity_pe}}
\end{figure}

\begin{figure}[t]
\includegraphics[scale=0.5]{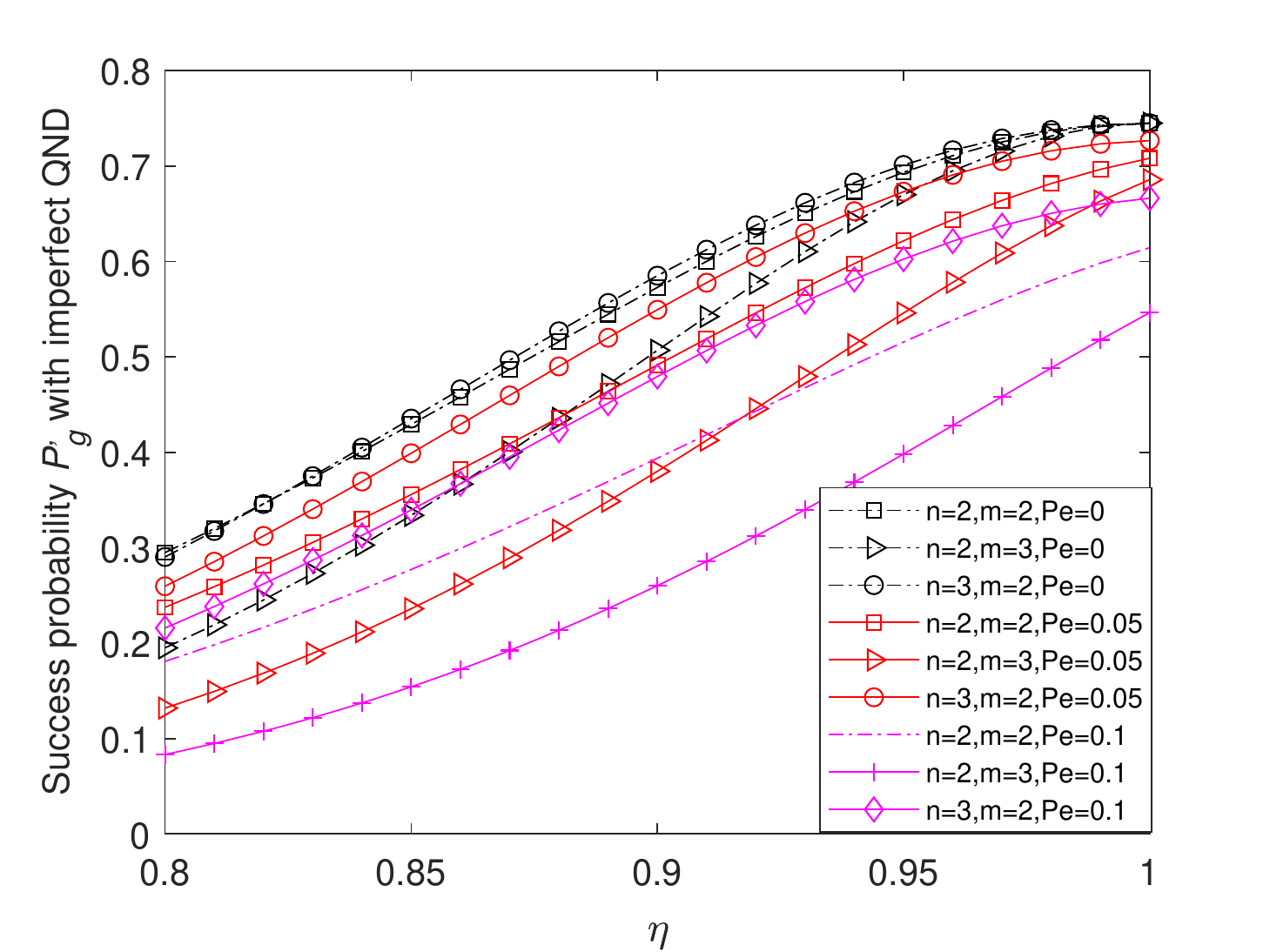}\caption{The success probability $P_g^{\prime}$ of the purification with the imperfect QND$_3$ versus $\eta$ with $F=0.85$ \cite{EPPPRAyan}.  \label{success_imper}}
\end{figure}

In practical scenario, the perfect QND$_3$ is hard to be available. Thus, the imperfection of QND$_3$ will cause errors occur on the BSA for the physical Bell states with the probability of $P_e$. According to Eq. (\ref{lowerlevel}) and the selection rule of this MBEPP, the fidelity can be given by
\begin{eqnarray}\label{LBSMImperfectFidelity}
{F_2} = \frac{{{F^2}P_4^2 + 4P_2^2P_3^2{{(1 - F)}^2} + 4{P_2}{P_3}{P_4}F(1 - F)}}{{{P_F}[P_4^2 + 4P_2^2P_3^2] + 8{P_2}{P_3}{P_4}F(1 - F)}}.
\end{eqnarray}
 Here ${P_4} = P_2^2 + P_3^2$, ${P_F} = {F^2} + {(1 - F)^2}$, $P_2=\sum\limits_{i \in even}^n {C_n^i}(1-P_1)^i(P_1)^{(n-i)}$, ${P_3} =\sum\limits_{j \in odd}^n {C_n^j}(1-P_1)^j(P_1)^{(n-j)}$, ${P_1} = \frac{{{{(1 - {P_e})}^m}}}{{P_e^m + {{(1 - {P_e})}^m}}}$ and $C_n^i = \frac{{n!}}{{i!(n - i)!}}$. Obviously, if $P_e=0$, it yields $F_2=F_1$. One can clearly see from Fig. \ref{Fidelity_pe} that the fidelity of the resultant state improves with the initial fidelity $F$, accordingly. In addition, an increasing $P_e$ will result in a reduction on fidelity $F_2$ after the MBEPP because the imperfect QND leads to ambiguously distinguish the parity of the Bell states in physical level. Moreover, the imperfection of QND$_3$ leads to the success probability of MBEPP for logical qubit rewrite as
\begin{eqnarray}\label{LBSMImperfectTotallSuccess}
P_g^{\prime} = \sum\limits_{j = 0}^{{n_t}} {\prod\limits_{s = 1}^4 {P_{{j_s}}^{\prime}} {E_j}} {(1 - \eta )^j}{\eta ^{4mn - j}}{P_{Fg}(n,m)},
\end{eqnarray}
with
\begin{eqnarray}\label{PFG}
{P_{Fg}(n,m)} \!=\! {{P_F}[P_4^2 + 4P_2^2P_3^2] \!+\! 8{P_2}{P_3}F(1 \!-\! F){P_4}},
\end{eqnarray}
which indicates that $P_g^{\prime} \le P_g$. Fig. \ref{success_imper} shows the adverse effect of imperfect QND$_3$ on the success probability of MBEPP for logical qubit entanglement. For instance, if we set $n=m=2$ and $\eta=0.8$, we have $P_g^{\prime}=0.18$ for $P_e=0.1$ and $P_g^{\prime}=0.3$ for $P_e=0$. Additionally, the success probability of $P_g^{\prime}$ accordingly increases with $\eta$. Moreover, it exists an optimal coding structure to maximize the success probability of the MBEPP for logical qubit entanglement when the same total number of photons is adopted.

\section{Discussion and Conclusion}
In this review, we have introduced some typical EPPs, including the basic entanglement purification theory, the EPPs with linear optics, the EPPs with cross-Kerr nonlinearity, the hyperentanglement EPPs, one deterministic EPP, and the MBEPPs. Some important progresses about entanglement purification experiments were also briefly introduced. Though many EPPs were proposed and the several experiments were also realized, it still remains a big challenge in practical applications. In the theoretical side, new entanglement purification protocols such as the MBEPP were proposed. These EPPs still require further investigation. In the experimental side, one can explore the experiments of EPPs in some other solid quantum systems such as ion traps and quantum dot systems. In addition, the entanglement of two atoms via fibres was distributed over dozens of kilometres \cite{atomentanglement}. Hence, the long-distance entanglement purification in solid state systems can be investigated in the future.  Moreover, the existing experiments realized the bit-flip error correction for one round. Actually, the fidelity of the mixed state can be improved after multi-step purifications. As a consequence, the researchers can pay attention to developing a high-efficient multi-step EPP to further improve the fidelity of the mixed state. Furthermore, one can devote to studying the combination of the entanglement purification, entanglement swapping, and entanglement memory in a meaningful distance to construct intact quantum repeaters. Finally, EPPs can be used in thr entanglement-based quantum communication such as QKD and QSDC to improve the secure key directly, as well as the device-independent QKD and device-independent QSDC to extend the distance in quantum communication.

\section*{ACKNOWLEDGEMENTS}
This work is supported by the National Natural Science Foundation of China under Grant  Nos. 11974189, 12175106, and the Natural Science Foundation of the Jiangsu Higher Education Institutions of China under Grant No. 20KJB140001.

\end{document}